\documentclass[aps,prb,twocolumn,superscriptaddress, longbibliography]{revtex4-2}

\usepackage{gensymb}
\usepackage[utf8]{inputenc}
\usepackage{amsmath,amssymb}
\usepackage{graphicx}
\usepackage{xcolor}
\usepackage{ulem}

\usepackage{bm,hyperref}
\usepackage[final]{changes}
\definechangesauthor[name = Georg, color=blue]{GK}
\definechangesauthor[name = G, color=green]{G}

\newcommand{\Tmaxc}{T^{\rm max}_{\rho_c}}
\newcommand{\Tmaxa}{T^{\rm max}_{\rho_a}}
\newcommand{\Tc}{T_{sc}}
\newcommand{\Hc}{H_{c2}}
\newcommand{\Tchimax}{T^{\rm max}_{\chi_b}}
\newcommand{\TWMO}{T_{\rm WMO}}
\newcommand{\TMO}{T_{\rm MO}}

\begin{document}

\title{\textbf{$\mathbf{c}$ axis electrical transport at the metamagnetic transition in the heavy-fermion superconductor UTe$_{2}$} under pressure}

\author{G. Knebel}
\affiliation{Univ. Grenoble Alpes, CEA, Grenoble INP, IRIG, PHELIQS, F-38000 Grenoble, France}
\author{A. Pourret}
\affiliation{Univ. Grenoble Alpes, CEA, Grenoble INP, IRIG, PHELIQS, F-38000 Grenoble, France}
\author{S. Rousseau}
\affiliation{Univ. Grenoble Alpes, CEA, Grenoble INP, IRIG, PHELIQS, F-38000 Grenoble, France}
\author{N. Marquardt}
\affiliation{Univ. Grenoble Alpes, CEA, Grenoble INP, IRIG, PHELIQS, F-38000 Grenoble, France}
\author{D. Braithwaite }
\affiliation{Univ. Grenoble Alpes, CEA, Grenoble INP, IRIG, PHELIQS, F-38000 Grenoble, France}
\author{F. Honda}
\affiliation{Institute for Materials Research, Tohoku University, Oarai, Ibaraki, 311-1313, Japan}
\affiliation{Central Institute of Radioisotope Science and Safety, Kyushu University, Fukuoka 819-0395, Japan}
\author{D. Aoki}
\affiliation{Univ. Grenoble Alpes, CEA, Grenoble INP, IRIG, PHELIQS, F-38000 Grenoble, France}
\affiliation{Institute for Materials Research, Tohoku University, Oarai, Ibaraki, 311-1313, Japan}
\author{G. Lapertot}
\affiliation{Univ. Grenoble Alpes, CEA, Grenoble INP, IRIG, PHELIQS, F-38000 Grenoble, France}
\author{W. Knafo}
\affiliation{Laboratoire National des Champs Magn\'etiques Intenses - EMFL, CNRS, Univ. Grenoble Alpes, INSA-T, Univ. Toulouse 3, F-31400 Toulouse, France}
\author{G. Seyfarth}
\affiliation{Laboratoire National des Champs Magn\'etiques Intenses (LNCMI), CNRS, Univ.~Grenoble Alpes, F-38042 Grenoble, France}
\author{J-P. Brison}
\affiliation{Univ. Grenoble Alpes, CEA, Grenoble INP, IRIG, PHELIQS, F-38000 Grenoble, France}
\author{J. Flouquet}
\affiliation{Univ. Grenoble Alpes, CEA, Grenoble INP, IRIG, PHELIQS, F-38000 Grenoble, France}

\email[E-mail me at: ]{georg.knebel@cea.fr}
\date{\today }

\begin{abstract}
The electrical resistivity of the unconventional superconductor UTe$_2$ shows very anisotropic behavior in the normal state depending on the current direction. In the present paper we show that the maximum in the resistivity $\rho_c$ for current applied along the $c$ axis at $\Tmaxc \approx 14.75$~K follows the minimum in the thermal expansion $T_\alpha^\star$ along $b$ axis. Under a magnetic field applied along the $b$ axis, $\Tmaxc$ can be tracked up to the critical point of the first order metamagnetic transition, which is located near 6~K and 34.5~T. Surprisingly, at the metamagnetic field $H_m$ the resistivity $\rho_c$ shows a steplike decrease while the resistivities $\rho_a$ and $\rho_b$, for current along the $a$ and $b$ axis, respectively, show a steplike increase. Under hydrostatic pressure $\Tmaxc$ and $H_m$ decrease significantly up to the critical pressure $p_c$ at which superconductivity is suppressed and a long range antiferromagnetic order appears. We show that the phase diagram at different pressures can be scaled by $\Tmaxc$ in field and temperature suggesting that this temperature scale is governing the main interactions in the normal state. 
\end{abstract}


\maketitle

\section{Introduction}

The discovery of superconductivity in UTe$_2$ has raised an intense research effort due to the possible spin-triplet pairing \cite{Ran2019, Aoki2019, Aoki2022review}.  Indication for this rare pairing state is the strong enhancement of the superconducting upper critical field $\Hc$, which  exceeds by far the Pauli limit of superconductivity for all three crystallographic axes. Another one is the change of this paramgnetic limit in the different superconducting phases observed under pressure for field along the easy a-axis \cite{Knebel2020}.  Knight shift measurements in magnetic resonance experiments (NMR) suggest a possible spin-triplet  superconducting state too \cite{Nakamine2021, Fujibayashi2022, Matsumura2023}. Moreover, a quite singular property of superconductivity in UTe$_2$ is that the pairing strength itself is sensitive to the magnetic field for all crystal axes \cite{Rosuel2023}.  For a field applied along the $b$ axis, superconductivity is enhanced above $H^\star \approx 17$~T \cite{Knebel2019, Ran2019a}, and a phase transition between two susperconducting phases, the low-field superconducting phase (lfSC) and a high-field phase (hfSC), have been evidenced by  specific heat \cite{Rosuel2023}, and independently by combined  NMR and ac susceptibility measurements \cite{Kinjo2023PRB}. This shows that the field-induced hfSC   is a bulk superconducting phase. 
It is  limited by a metamagnetic transition at $\mu_0 H_m \approx 35$~T  \cite{Knebel2019, Ran2019a}. The first order metamagnetic transition is characterized by a large jump of the magnetization \cite{Miyake2019}, a volume discontinuity \cite{Miyake2022}, and a jump in the magnetoresistance with the current applied along the $a$ axis \cite{Knafo2019}. It is accompanied by a strong increase of the Sommerfeld coefficient $\gamma$ of the specific heat \cite{Imajo2019, Miyake2021b, Rosuel2023} and  of the $A$ coefficent of the Fermi-liquid resistivity \cite{Knafo2019, Thebault2022}. Recent NMR measurements show that critical longitudinal magnetic fluctuations starts to develop for $H > H^\star$ and are diverging in the vicinity of $H_m$ \cite{Tokunaga2023}. 
 An abrupt Fermi-surface change at $H_m$ has been discussed from Hall effect measurements \cite{Niu2020b}, but constitutes still an open question \cite{Helm2024}. Superconductivity in UTe$_2$ was claimed to be topologically non-trivial following the detection of chiral-edge states in scanning-tunneling microscopy (STM) \cite{Jiao2020}, and of broken time-reversal symmetry in Kerr effect studies\cite{Hayes2021}. However, these results are still discussed controversally and may depend on the sample quality \cite{Azari2023, Ajeesh2023}.

Under hydrostatic pressure, two superconducting phases occur above 0.3~GPa, indicating multiphase superconductivity at zero field \cite{Braithwaite2019, Thomas2021, Lin2020}.  Different superconducting phases have been confirmed by measurements under magnetic field applied along the $a$ and $b$ axes at constant pressure \cite{Aoki2020, Lin2020}. The superconducting phases are suppressed at a critical pressure $p_c \approx 1.5$--1.7~GPa (depending on the pressure medium) and a magnetically ordered state occurs. Initially a ferromagnetic state has been expected \cite{Ran2020}, but recent neutron diffraction experiments under confirm an incommensurate antiferromagnetic order above $p_c$ \cite{Knafo2023}. The magnetic anisotropy is reversed at $p_c$ \cite{Li2021, Thomas2020, Valiska2021, Kinjo2022PRB}. 
While the $a$ axis is the easy magnetization axis at ambient pressure, it switches to the $b$ axis in the pressure induced magnetic phase. The change in the magnetic anisotropy has also a feedback on the superconducting state under pressure. Close to $p_c$, the upper critical field $H_{c2}$ is the highest for the $c$ axis, and even reentrant superconductivity occurs in this direction \cite{Knebel2020, Ran2020, Aoki2021, Valiska2021}. 

To understand the occurrence of these various superconducting phases in UTe$_2$ the knowledge of the electronic structure and of the magnetic fluctuations is necessary. Various electronic band structure calculations indicate cylindrical Fermi surfaces along the $c$ axis \cite{Xu2019, IshizukaPRL2019, Fujimori2019, Miao2020, Shishidou2021, Choi2022}. Experimentally the Fermi surface of UTe$_2$ is not fully determined, but angle resolved photoemission spectroscopy (ARPES) experiments \cite{Miao2020} and quantum oscillations have confirmed the quasi-two-dimensional Fermi-surface sheets \cite{Aoki2022dHvA, Eaton2024, Broyles2023, Aoki2023note}. 

Resistivity measurements with different current directions suggest that UTe$_2$ is a three dimensional metal \cite{Eo2022} indicating that either the cylinders are strongly corrugated or that a three-dimensional Fermi surface is missing in quantum oscillation experiments. For currents applied along the $a$ and $b$ axes, the temperature dependence of the resistivity shows a broad maximum near 60~K and drops for lower temperatures, indicating the formation of coherent quasiparticle bands characteristic of a heavy-fermion state. Once a non-magnetic background of the scattering is subtracted, this maximum is shifted down to the temperature scale $\Tchimax \approx 35$~K,  for which the susceptibility and the Hall effect for $H\parallel b$ are maximum \cite{Valiska2021, Thebault2022, Niu2020b}.  For a current applied along the $c$ axis a distinct maximum of the resistivity appears only at $\Tmaxc \approx 14.5$~K \cite{Eo2022, Thebault2022, Girod2022, Kim2023}. Near this characteristic temperature several other quantities show anomalies: the electronic specific heat has a broad maximum \cite{Willa2021}, the thermal expansion coefficients along the $c$ and $b$ axes have a minimum \cite{Willa2021, Thomas2021}, the thermoelectric power has a minimum \cite{Niu2020}, and the $a$ axis susceptibility $\chi_a$ as well as the Knight shift becomes constant at lower temperatures \cite{Tokunaga2022}. The anomaly at $T^\star$ corresponds to the crossover to a coherent low temperature regime \cite{Eo2022, Butch2022}. Furthermore, inelastic neutron scattering experiments show the development of a magnetic excitation below $T \approx 60$~K at the incommensurate wavevector $\mathbf{k_1} = (0, 0.57, 0)$ which is maximal at the energy transfer 3-4 meV \cite{Duan2020, Knafo2021, Butch2022}. They also reveal that these fluctuations are  low dimensional and antiferromagnetic, saturating below $T^\star \approx 15$~K \cite{Duan2020, Knafo2021}. They become gapped in the superconducting state \cite{Duan2021, Raymond2021}.

In the present paper, we study the field and pressure dependence of the different fluctuations by electrical resistivity measurements. The paper is divided in different sections. In Section \ref{experimental_methods} we will present the experimental details and give a summary of all samples studied in this work. Section \ref{ambient_pressure} presents ambient pressure results: we will show the field dependence ($H \parallel b$) of the $T^\star$ anomaly in the resistivity with current applied along the $c$ axis ($\rho_c$) and longitudinal thermal expansion measurements ($\alpha_b$) up to the metamagnetic field. We will demonstrate that the maximum $\rho_c$ at $T^\star$ can be followed up to the metamagnetic transition. Further we will discuss the anisotropy of the magnetoresistivity for currents applied along the $a$, $b$, and $c$ axes. Finally, we will compare in the superconducting phase diagram in a field $H \parallel b$ for different samples with different $\Tc$. We find that the lfSC phase scales with $\Tc$, while the hfSC is strongly affected by sample quality.

In the following Section \ref{resistivity_pressure} we will concentrate on the resistivity measurements under high pressure. In Section \ref{pressure_PD} the temperature dependence of the resitivity under pressure for a current applied along the $c$ axis are presented and compared to those with a current along the $a$ axis. We present the pressure-temperature phase diagram of UTe$_2$ at zero magnetic fields. In Section \ref{resistivity_P1-P5} we present the temperature and field dependence of $\rho_c$ under pressure. In Section \ref{discussion} we discuss the obtained magnetic field -- temperature phase diagrams for the different pressures. We show that the phase diagrams for the different pressures scale with the temperature $T^\star$ indicating the importance of the fluctuations connected with this energy scale. Magnetoresistivity gives evidence that the order, which occurs above the critical pressure, is antiferromagnetic in agreement with recent neutron diffraction results \cite{Knafo2023}. Finally in Section \ref{conclusion} we give a short conclusion. Additional data supporting the analyses are presented in the Appendix \ref{supplementary_data}.

\section{Experimental methods}
\label{experimental_methods}
 
In this study single crystals of UTe$_2$ grown by chemical vapor transport (CVT) using iodine as transport agent \cite{Aoki2020SCES} and by the molten salt flux  (MSF) method \cite{Sakai2022} have been investigated. An overview of all samples is given in Tab.~\ref{table_crystal}. At ambient pressure we studied the dependence of the current direction on the electrical resistivity and on the $A$ coefficient of the Fermi-liquid resistivity. 
For $J \parallel b$ (sample S4) and $J \parallel c$ (S1a, S5) single crystals grown by CVT have been investigated with  superconducting transitions $T_{sc} \approx 1.85$~K  at zero field  (throughout the paper, the superconducting transition temperature $T_{sc}$ in the resistivity is defined by the criteria $\rho = 0$, see Fig.~\ref{def_Tsc_Tstar} in the Appendix). CVT grown samples have been cut from large crystals with a spark cutter after alignment using a Laue photograph.  Typical dimensions of these CVT crystals are $1\times0.3\times0.15$~mm$^3$. Two different crystals have been used for measurements with current $J \parallel c$. For sample S1a, the alignment of the contacts has probably not been perfectly done due to its irregular, non bare-like shape, and the absolute value of the resistivity of this sample is smaller than expected from literature \cite{Eo2022, Girod2022}. We studied in addition a second  sample S5 ($0.8\times0.15\times0.1$~mm$^3$) which has a similarly high resistance; unfortunately, here  the field orientation is not perfect as the measured metamagnetic transition in this experiment is higher than expected (data are shown in the Appendix Fig.~\ref{Annex_rho_c}). For resistivity measurements with current $J \parallel a$ axis ($\rho_a$) we used a MSF sample with a $T_{sc} \approx 2$~K. The MSF sample had a naturally needlelike shape along the $a$ axis with dimensions of $2\times0.2\times0.2$~mm$^3$. The high-pressure resistivity experiments with current alonc $c$ have been performed on a bareshape-like sample (S1b) which is cut from the same single crystal as S1a with current applied along the $c$ axis. For the measurements with current along $a$ we refer to our previous results reported partly in Ref.~\cite{Knebel2020}.

\begin{table}[b]
\caption{Summary of single crystals used in this study and the corresponding measurements.}
\label{table_crystal} 
	\begin{tabular}{cccc}
		\hline
		\hline
			 & crystal rowth & $\Tc$ (K) & measurement \\
		\hline
		S1a	& CVT & 1.84 & $\rho_c (T, H)$, $J \parallel \sim c$, $p = 0$ \\
			&	&	& (current not perfectly aligned) \\
		S1b	& CVT & 1.84 & $\rho_c (T, H)$, $J \parallel c$, $p \neq 0$ \\
		S2	& CVT & 1.45 & $\alpha_b$, same crystal as in \cite{Rosuel2023} \\
		S3	& MSF & 2.0 & $\rho_a (H)$, $J\parallel a$,  $p = 0$ \\
		S4	& CVT & 1.82 & $\rho_b (H)$, $J\parallel b$,  $p = 0$ \\
		S5  & CVT & 1.85 & $\rho_c (H)$, $J\parallel c$, $p = 0$ \\
		& & & (not perfectly aligned in field) \\
		\hline
		\hline
	\end{tabular}

\end{table}

Electrical resistivity measurements have been performed with a standard four point lock-in technique with a maximal applied current of 0.5~mA for the measurements at ambient pressure and 1~mA in the high pressure cell. For all samples, electrical contacts to the samples have been performed by spot welding 15 $\mu$m diameter Au wires on a freshly polished surface of the crystal as it is known that the surface of UTe$_2$ is very sensitive to air. To strengthen the electrical contacts mechanically, tiny drops of conducting silver paste have been added on top of the welded contacts. The contact resistance is typically of the order of several m$\Omega$, but this has not been further studied.

The high-field transport measurements have been performed at the high magnetic field laboratory LNCMI in Grenoble using the resistive 36~T magnet M9 using a $^3$He cryostat with a base temperature near 400~mK. We used a  MP35N piston cylinder pressure cell to apply pressures up to 1.61 GPa with Daphne 7373 oil as pressure medium for the high-pressure resistivity experiment. The outer diameter of the pressure cell is only 15~mm which fits very close to the inner diameter of the vaccuum can of the cryostat (16 mm). The high-pressure experiments have been performed up to a maximal field of 35~T. The temperature has been measured with a calibrated RuO$_2$ chip which is glued on the pressure cell.  The lowest temperature in the pressure cell was $T_{\rm min} \approx 1.2$~K. 

Thermal expansion measurements at zero pressure have been performed using a high resolution capacitive dilatometer \cite{Kuechler2012} in the LNCMI Grenoble on the magnet M10 with a maximal field of 30~T in the temperature range from 2 to 25~K. The CVT grown sample (S2) with $T_{sc} \approx 1.5$~K was already studied in Ref.~\cite{Rosuel2023} by magnetostriction experiments. 

\section{Ambient pressure results}
\label{ambient_pressure}

\subsection{c axis transport and field dependence of $\mathbf{T^\star}$}

\begin{figure}[bt]
	\includegraphics[width=0.9\linewidth]{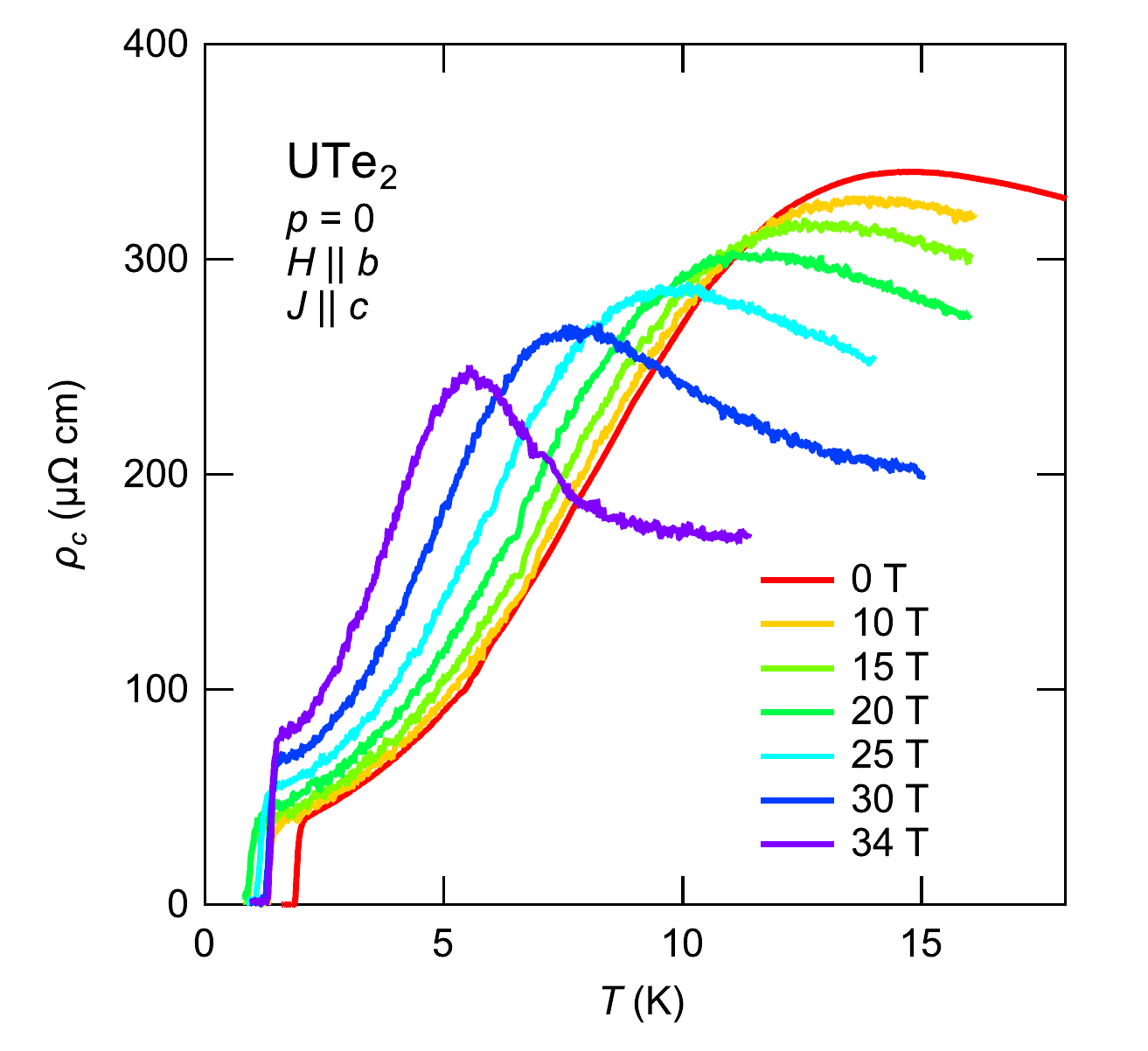}
	\caption{Temperature dependence of the electrical resistivity $\rho_c (T)$ of UTe$_2$ with current $J \parallel c$ at different magnetic fields $H \parallel b$ (measured on sample S1a). The maximum of the resistivity $T^{\rm max}_{\rho_c}$ shifts with field from $T^{\rm max}_{\rho_c} = 14.75$~K at $H=0$ to 5.5~K at 34~T.}
	\label{Fig01}
\end{figure}

Figure \ref{Fig01} displays the temperature dependence of the $c$ axis resistivity $\rho_c$ for different magnetic fields applied along the $b$ axis below 20~K. A maximum of the resistivity at $T^{\rm max}_{\rho_c} \approx 14.75$~K occurs at $H = 0$, similar to $\rho_c$ measurements \cite{Eo2022, Thebault2022}. However, as already mentioned above, the absolute value of the resistivity $\rho_c$ at low temperature is smaller than in previous reports \cite{Eo2022, Girod2022, Thebault2022}, which might be due to a non-perfect alignment of the current with respect to the $c$ axis.   Under magnetic field $\Tmaxc$ shifts to a lower temperature and the maximum gets more pronounced. At the highest field of 34~T, just below the metamagnetic transition, $\Tmaxc \approx 5.5$~K. It is close to the critical point of the first order metamagnetic transition which is located near 6~K and 34.5~T. Superconductivity, defined by $\rho = 0$, occurs in zero field below $T_{sc}= 1.89$~K. The minimum of $T_{sc}$ ($T_{sc} = 0.88$~K) is found at $H^\star \approx 18$~T. For higher fields the temperature of zero resistivity increases again and at 34~T, $T_{sc} = 1.35$~K. This behavior of $H_{c2} (T)$ is similar to that of previous reports \cite{Knebel2019, Ran2019a, Sakai2023}. The highest $\Tc$ of the hfSC phase  is found just below $H_m$. 

 In the normal state above $T_{sc}$ the resistivity $\rho_c$ follows a Fermi-liquid temperature dependence $\rho (T) = \rho_0 + A T^2$, where $\rho_0$ is the residual resistivity and $A$ the coefficient of the electron-electron scattering term. This is shown in Fig.~\ref{rho_vs_T2} in the Appendix, which shows $\rho_c$ as a function of $T^2$ for several magnetic fields \cite{footnote1}.
 While at 10~T the fitting range is up to about 6~K, for the highest field this range is strongly reduced and the maximum temperature of the fit is about 4~K, i.e.~the fitting range of the $T^2$ dependence is small so that the following analysis of the $A$ coefficient is rather qualitative. Anyway, in heavy-fermion systems the relation $A \propto \gamma ^2$ is often obeyed indicating  that local fluctuations dominate \cite{Jacko2009}. In a simplified picture, the $A$ coefficient is an indirect measure the square of the effective mass $m^*$ of the charge carriers. One has to keep in mind that resistivity in difference to specific heat is a directional probe and so in addition to the scattering time it depends also on the topology of the Fermi surface and not only on the density of states. The field dependence of $A$ and the residual resistivity $\rho_0$ will be discussed below.

\begin{figure}[t]
	\includegraphics[width=0.9\linewidth]{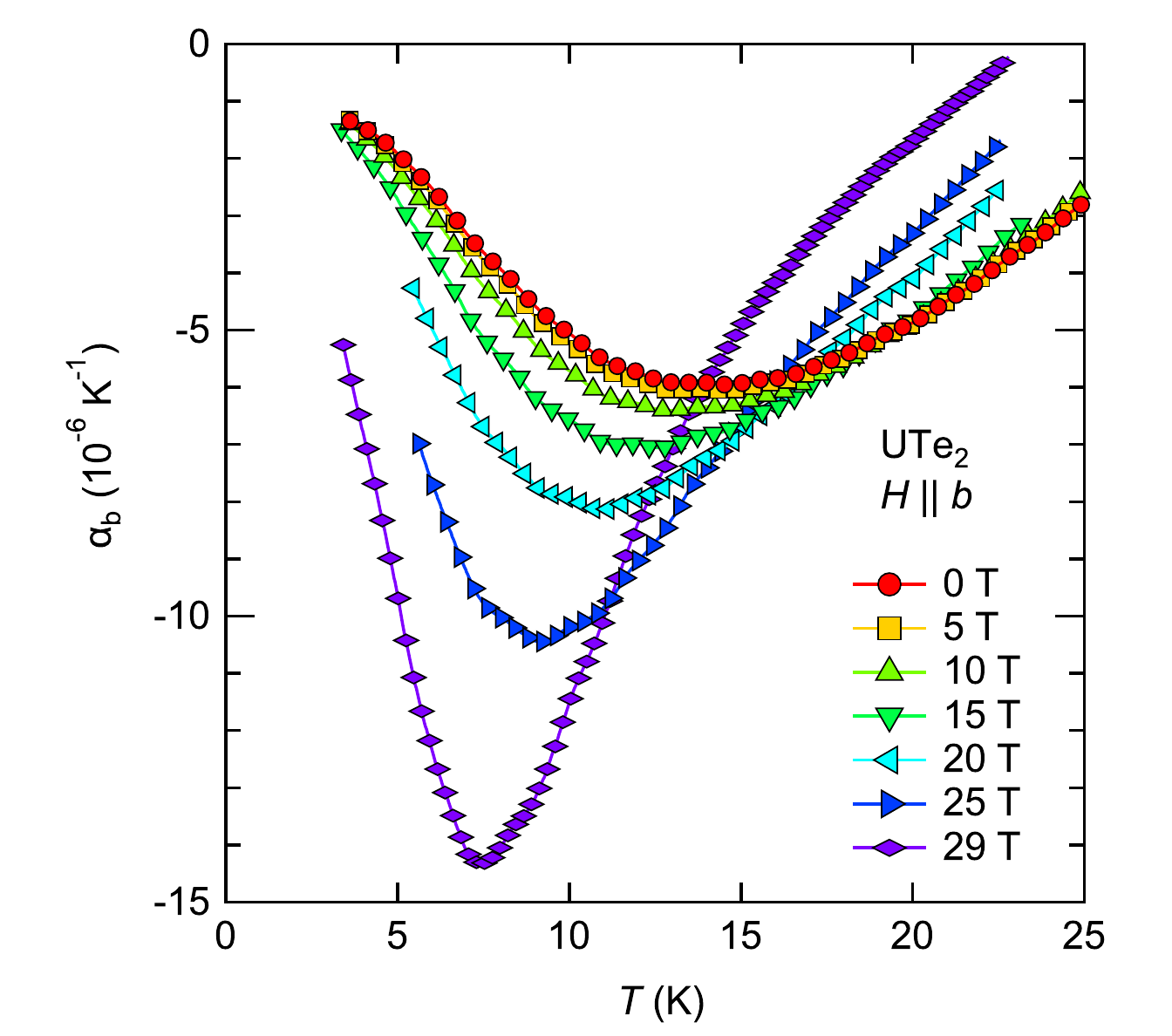}
	\caption{Longitudinal thermal expansion $\alpha_b$ as a function of temperature for different magnetic fields applied along $b$ axis up to 29~T.  $\alpha_b (T)$ shows a pronounced minimum, which we refer as $T^\star_\alpha$ in the text. Measurements are performed on sample S2. }
	\label{Fig02}
\end{figure}

\begin{figure}[t]
	\includegraphics[width=0.9\linewidth]{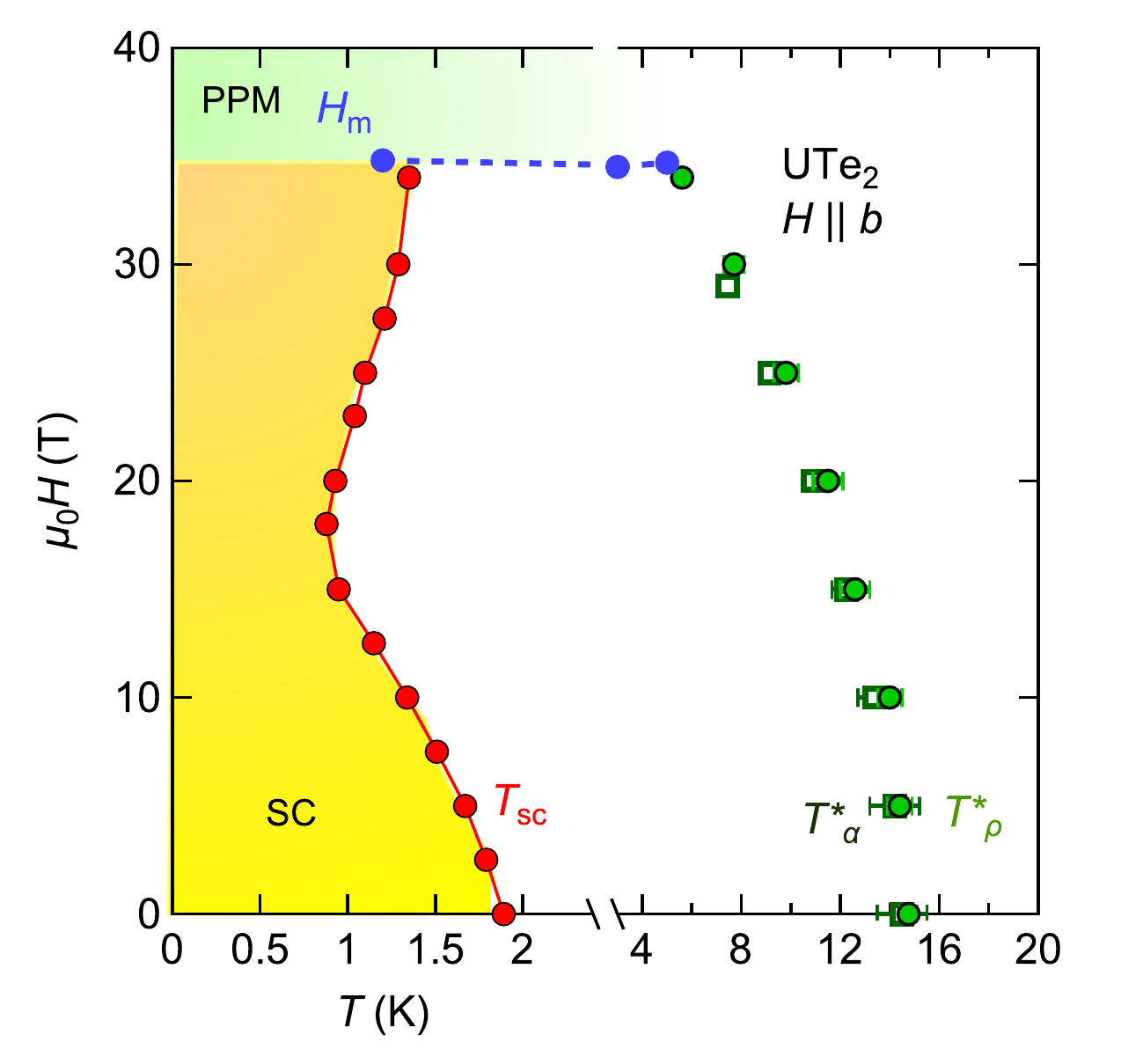}
	\caption{Phase diagram of UTe$_2$ for field along the $b$-axis from the $c$ axis transport and the thermal expansion measurements. $\Tc$ is the superconducting transition temperature, $T^\star_\alpha$ the minimum in the thermal expansion coefficient $\alpha_b$, $\Tmaxc$ the maximum in $\rho_c(T)$, and $H_m$ the field of the metamagnetic transition. Clearly, the maximum $\Tmaxc$ coincides with the temperature of the minimum in the thermal expansion. $\Tc$ is determined by the criterium $\rho = 0$ and error bars are smaller than the symbol size.}
	\label{Fig03}
\end{figure}

Figure \ref{Fig02} shows the longitudinal thermal expansion coefficient along the $b$ axis  $\alpha_b$ as a function of temperature at different magnetic fields. $\alpha_b (T)$ is negative and has a very broad minimum at $T^\star_\alpha \approx 14.5$~K in zero field.  A magnetic field applied along the $b$ axis shifts $T^\star_\alpha$ to lower temperatures and the minimum sharpens significantly. We can follow the minimum up to the highest field of 29~T and $T^\star_\alpha (H)$ determined from thermal expansion is in excellent agreement with the maximum $\Tmaxc$ in the $c$ axis transport. Fig.~\ref{Fig03} shows the phase diagram of UTe$_2$ at ambient pressure determined from the $c$ axis transport and the thermal expansion along $b$ axis. This shows that the thermodynamic anomaly detected at $T^\star_\alpha$ connects to the metamagnetic transition at $H_m$. The $T^\star_\alpha \approx \Tmaxc$ is a signature of a crossover to the low temperature state. In heavy fermion compounds this crossover arises from  the interplay of the onsite and intersite magnetic interactions, which result in the formation of the coherent heavy-fermion state on cooling.  In UTe$_2$ this crossover is also visible in the evolution of the magnetic fluctuations at finite wave vectors \cite{Knafo2021, Duan2020, Butch2022, Tokunaga2022, Eo2022}. In addition to the magnetic fluctuations, crystal field effects may also play a role in the microscopic origin of the  $T^\star$ anomaly \cite{Khmelevskyi2023}. 

\subsection{Anisotropic transport at $H_m$}

\begin{figure}[t]
	\includegraphics[width=0.9\linewidth]{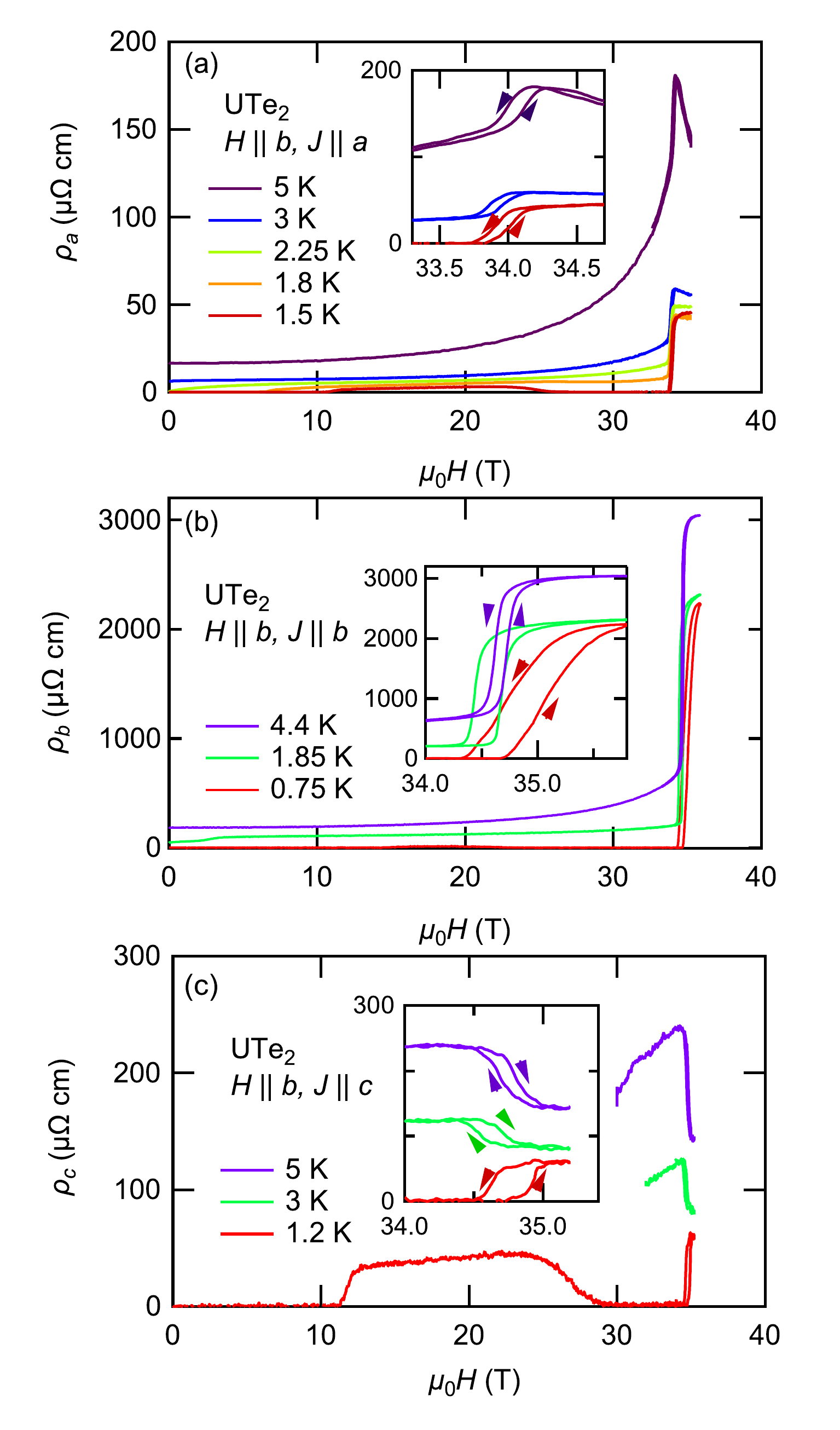} 
	\caption{Magnetoresistivity for $H \parallel b$ at different temperature with the current applied along $a$, $b$ and $c$ axes. The inset in every panel gives the field range of the metamagnetic transition in an enlarged scale. [MSF-grown sample S3 for $J || a$, CVT grown samples for $J || b$ (S4) and $J || c$ (S1a).]  }
	\label{magnetoresistance_all}
\end{figure}

Previous magnetoresistivity measurements with a current  $J \parallel a$ axis showed that $\rho_a (H)$ in the normal state just above $T_{sc}$ increases abruptly at the metamagnetic transition $H_m$ by a factor of 4 \cite{Knafo2019, Knafo2020, Niu2020b}. A hysteresis between field up and down sweeps of about 0.2~T indicates the first order nature of the transition. The transition is sharp up to a critical point of $\approx 6$~K where the the hysteresis vanishes. At higher temperatures a broad crossover from a coherent correlated paramagnetic to an incoherent paramagnetic state occurs. A similar abrupt change in the Hall resistivity as a function of field at $H_m$ together with a change of the main charge carriers observed in thermoelectric power experiments \cite{Niu2020b} suggest that an abrupt change of the Fermi surface occurs at the metamagnetic transition $H_m$.   
Figure \ref{magnetoresistance_all} displays the magnetoresistivity with a current injection along the $a$, $b$, and $c$ axes as a function of the magnetic field applied along the $b$ axis. For the $J \parallel a$ axis, we used a high quality MSF grown sample with a $\Tc = 2$~K.  $\rho_a (H)$ of the MSF sample with the higher $\Tc$  shows a similar field dependence in the normal state as reported previously \cite{Knafo2019, Knafo2020}
The samples used for the different current directions show re-entrant superconductivity at the lowest temperature up to $H_m$. 

For a current $J \parallel b$  axis, $\rho_b$ shows an extremely large positive jump at $H_m$, by a factor 8.5 at 2.25~K in the normal state, see Fig.~\ref{magnetoresistance_all}(b). This jump is much stronger than that being observed for a current applied along the $a$ axis. 

Astonishingly, for a current $J \parallel c$ axis, the magnetoresistivity $\rho_c(H)$ drops at the metamagnetic transition as shown in the lower panel of Fig.~\ref{magnetoresistance_all}(c); at 3~K the magnetoresistivity decreases by a factor of 1.45, i.~e., the change of the magnetoresistivity at the metamagnetic transition is much smaller and opposite to that observed along the  $a$ and $b$ directions, and for $T$ near $\Tc$ no strong anomaly occurs near $H_m$. With increasing temperatures above $\Tc$ the drop of the resistivity along $c$ becomes stronger and changes into a crossover above the critical point of the metamagnetic transition line, like for the other current directions (see also Ref.~\cite{Thebault2022}). These distinct differences between the magnetoresistivity measured for the different current directions might be due to the anisotropic quasi-2D Fermi surface of UTe$_2$ \cite{Aoki2022dHvA, Eaton2024, Broyles2023}, but a detailed microscopic picture of the anisotropic scattering is still missing. In Ref.~\cite{Eo2022}, within a simple two band model, it has been proposed that the conduction along the $c$ axis is dominated by a heavy $Z$ pocket of the Fermi surface detected in ARPES measurements, while the conduction along the $a$ and $b$-axis directions is dominated mainly by the cylindrical Fermi surfaces. Recent dHvA quantum oscillations experiments did not detect such a small 3D Fermi-surface pocket \cite{Aoki2022dHvA, Eaton2024, Aoki2023note} in difference to  Ref.~\cite{Broyles2023} where a 3D Fermi surface pocket may be observed using a tunnel diode oscillator circuit (TDO); future experimental investigations have to clarify this point.

An anisotropy of the transport properties at a metamagnetic transition has been also observed in other heavy-fermion compounds.  In the paramagnetic and nearly antiferromagnetic  CeRu$_2$Si$_2$, where the metamagnetic transition occurs for a field applied along the easy $c$ axis, the magnetoresitivity shows a positive jump in the transversal configutation $J \parallel a$, while a peak occurs for $J \parallel c$ \cite{Boukahil2014}. In the paramagnetic and nearly ferromagnetic UCoAl the magnetoresistance has a positive jump at $\mu_0 H_m \approx 0.6$~T for $J \parallel H \parallel c$, which is the easy magnetization axis, but a negative jump occurs in the transversale configuration  \cite{Matsuda2000}. And in antiferromagnetic UPd$_2$Al$_3$, a sharp peak occurs at $H_m$ in the longitudinal configuration while a strong decrease of almost 50 \% occurs in the transverse magnetoresistance \cite{deVisser1993UPd2Al3}. In all of these examples, the metamagnetic transition is accompanied with a Fermi surface reconstruction \cite{HAoki2014, Palacio-Morales2013, Terashima1997}, as proposed for UTe$_2$. 
\begin{figure}
	\includegraphics[width=1\linewidth]{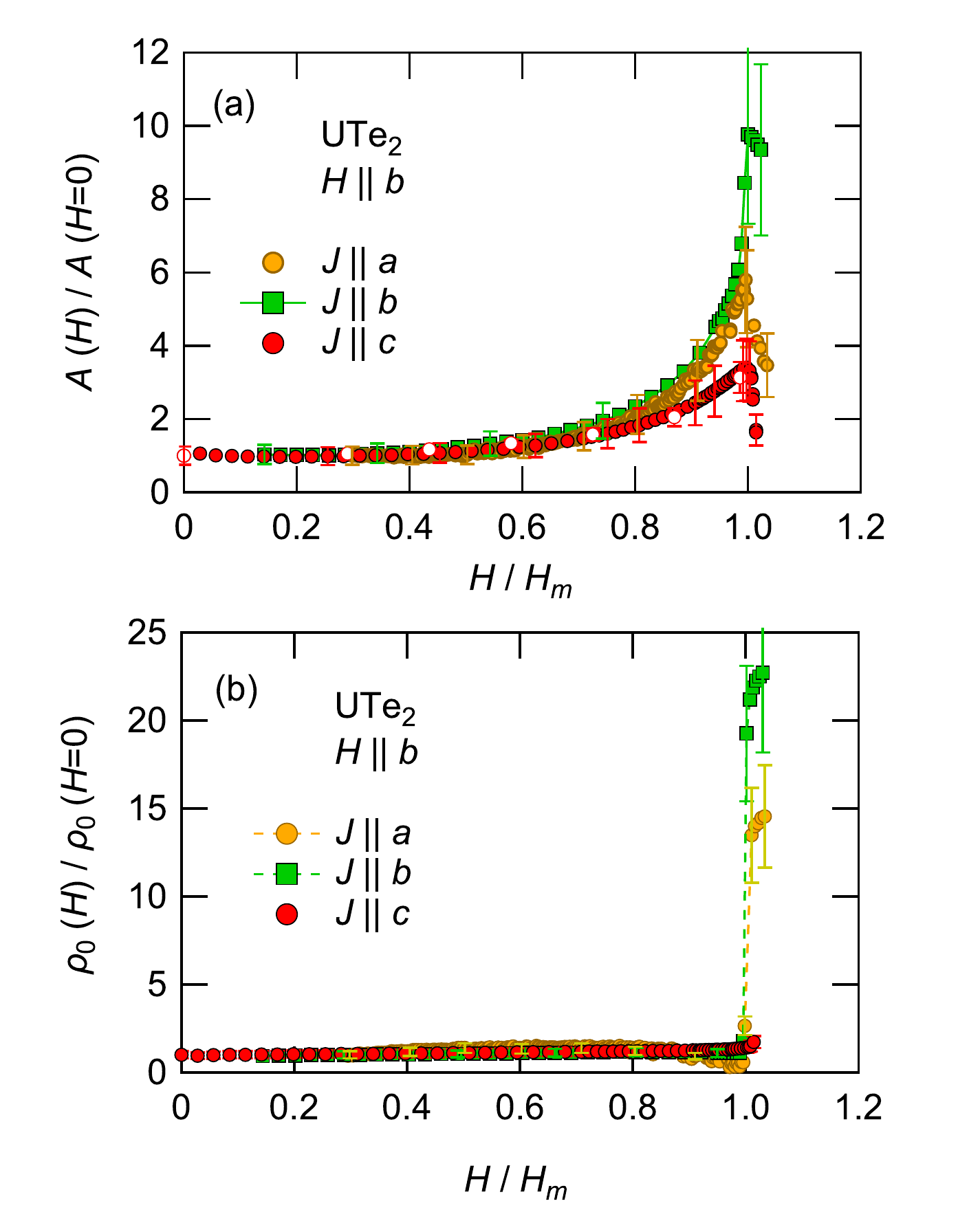}
	\caption{Normalized $A$ coefficient determined from a Fermi liquid temperature dependence $\rho (T) = \rho_0 + AT^2$ of the resistivity for current applied along the $a$, $b$, and  $c$ directions as a function of $H/H_m$. Full symbols are determined from the measured field sweeps at different temperatures. Fits have been performed for $T<3.5$~K $J || a$ (sample S3), $T<4.5$~K for $J || b$ (S4),, and $T<4$~K for  $J || c$ (S5, data shown in Fig~\ref{Annex_rho_c}). Open symbols for $J \parallel c$ are from fits of data shown in Fig.~\ref{Fig01} (sample S1a). The lower panel (b) shows the residual resistivity $\rho_0$ obtained from the Fermi liquid fit for the three current directions.}
	\label{Fig05}
\end{figure}

 The field dependences of the normalized $A$ coefficient and of the residual resistivity $\rho_0$ extracted from Fermi-liquid fits are shown in Fig.~\ref{Fig05} as a function of $H/H_m$. $A$ and $\rho_0$ were determined from the temperature dependent measurements (open circles for $J \parallel c$) and from the field sweeps at constant temperature (full symbols), see also Figs.~\ref{Annex_rho_c}, \ref{rho_vs_T2}, \ref{Annex_rho_Jb_and_Ja} in the Appendix. We compare $A (H)$ \added{normalized to its zero field value} for the different current directions in Fig.~\ref{Fig05} (\added{see also Fig.~\ref{Annex_A_coef_all_directions_absolute} in the Appendix which shows the absolute values}). The anisotropy of the $A$ coefficient at zero field is similar to that in Ref.~\cite{Eo2022}. The field dependence of $A$ for $J\parallel a$ determined on the new MCF sample is in excellent agreement with those published in Ref.~\cite{Knafo2019}. However, there are distinct differences in the field dependence of $A(H)$ for the different directions.  
 
 For $J \parallel a$ axis the absolute value of $A$ is the smallest and it increases from about $A =0.6\,\mu\Omega \,\rm{cm}\, K^{-2}$ at zero field to  4~$\mu\Omega \,\rm{cm}\, K^{-2}$ at $H_m$. Astonishingly, the field enhancement of $A$ is symmetric around $H_m$, which was not expected owing to the first order character of the metamagnetic transition. 
 For current along the $b$ axis, the absolute value of $A$ equals $5\, \mu\Omega \,\rm{cm}\, K^{-2}$ at zero field.  $A(H)$ increases very strongly by a factor of 10, when the metamagnetic transition is approached. (The absolute value of $A (H)$ for the $b$-axis direction may be over estimated as it is only determined from the measured field dependences at different temperatures, and no temperature dependent measurements at fixed field have been performed (see Fig.~\ref{Annex_rho_Jb_and_Ja}).)
 
 In zero field, the value of $A$ is the highest for a current applied along the $c$ axis, with $A = 8.3\, \mu\Omega \,\rm{cm}\, K^{-2}$. On approaching the metamagnetic transition under magnetic field, $A$ increases only by a factor of 3.4 at $H_m$ for $J \parallel c$ and decreases abruply just above the metamagnetic transition. If we compare the value at zero field with Sommerfeld coefficient at zero field of 120~mJ mol$^{-1}$K$^{-2}$ we find that $A/\gamma^2$ is much higher than the standard value of the Kadowaki Woods ratio of $1 \times 10^{-5}$~$\mu\Omega$ cm(K mol/mJ)$^2$, commonly expected in a Kondo lattice. The main reason for this strong difference is the quasi-2D Fermi surface in UTe$_2$, and thus, leading to a strongly anisotropic $A$ coefficient. By contrast, if we compare the field dependence of the specific heat divided by temperature $C/T$ at 1.8~K and that of the $A$ coefficient determined from $\rho_c$ (see Fig.~\ref{Annex_A_gamma} in the Appendix), we find an excellent agreement of both properties. Especially, the strong drop of the $C/T|_{T=1.8\,K}$ does not occur in the $A(H)$ for the other directions.  
  
 Figure \ref{Fig05}(b) shows the field dependence of the residual resistivity $\rho_0$ normalized to its value in zero field. We clearly see that $\rho_0$ for the $a$ and $b$ axes increases by a factor of at least 15 ($a$) or 20 ($b$), while the change of $\rho_0$ along $c$ is very small.  The strong increase of the residual resistivity for $J \parallel a$ and $b$ cannot be only due to a change in the charge carrier density, as this should influence the change of the magnetoresistance for $J \parallel c$ at $H_m$ in rather similar way.  In a simple picture it may also be an indication for a Fermi surface change at $H_m$ as has been previously suggested from the thermoelectric response through $H_m$ \cite{Niu2020b}. Surely, the change of the Fermi surface above $H_m$ is still an open question.

\subsection{Comparison of the superconducting phase diagrams}

\begin{figure}
	\centering
	\includegraphics[width=0.8\linewidth]{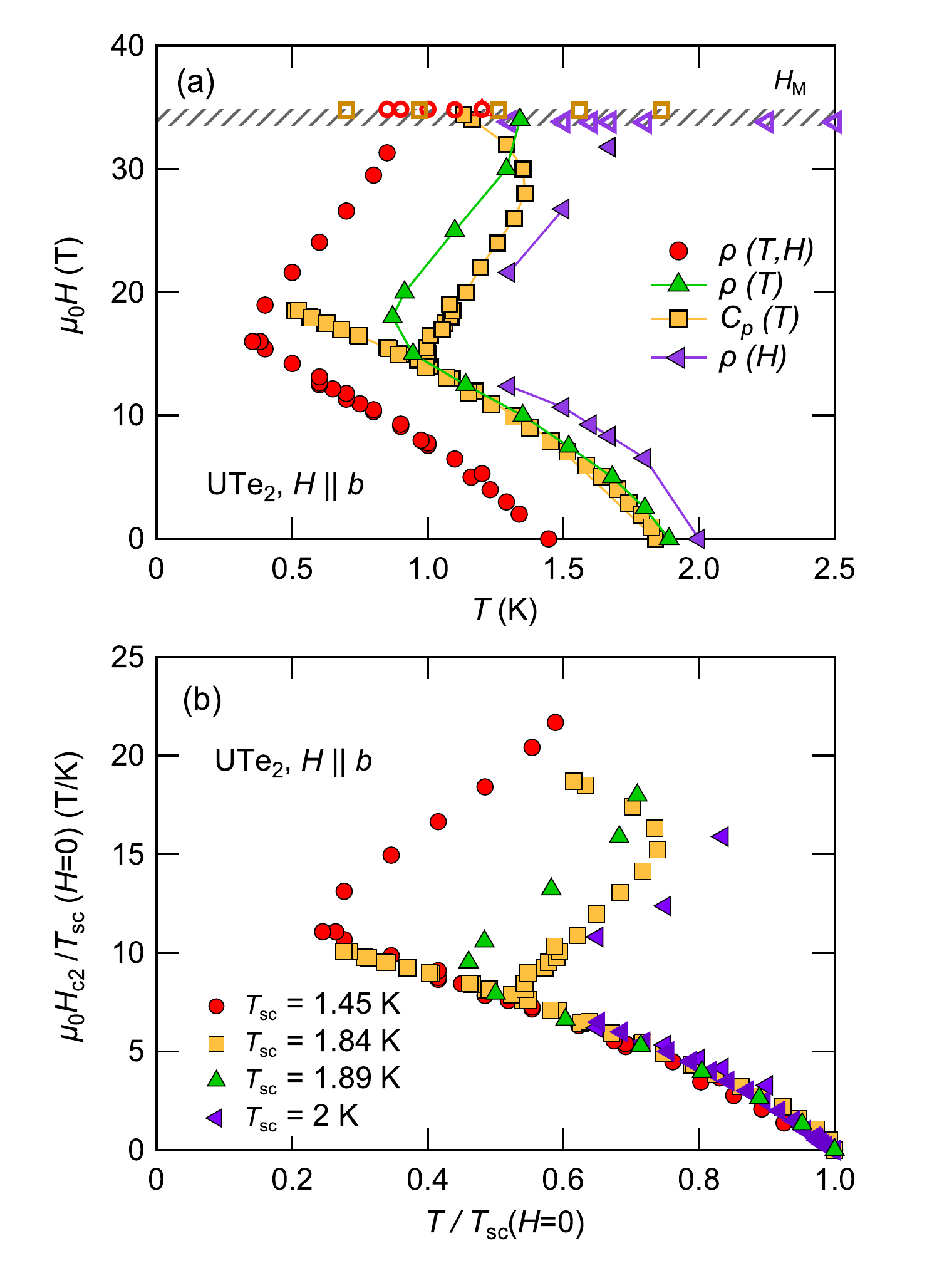}
 	\caption{(a) Comparison of the superconducting phase diagram of samples with different $\Tc$ for field $H\parallel b$. (Red circles are taken from Ref.~\cite{Knebel2019},  yellow squares from specific heat measurements of Ref.~\cite{Rosuel2023}, green triangles up determined from data in Fig.~\ref{Fig01},  violet triangles from Fig.~\ref{Fig03}(a).) The field of the metamagnetic transition is marked by open symbols, respectively. (b) Upper critical field normalized by $\Tc$ versus  $T/\Tc$ for the different samples.   }
	\label{comparisonpd}
\end{figure}

In Fig.~\ref{comparisonpd}(a) we compare the superconducting upper critical field $\Hc$ with $H \parallel b$ for samples with different values of $\Tc$ varying from 1.45~K to 2~K. All samples show field-reentrant superconductivity in fields above $H^\star \approx 16-18$~T.  The thermodynamic phase diagram has been only determined in one of the samples ($\Tc = 1.84$~K by specific heat measurements \cite{Rosuel2023}). The exact field of the reinforcement superconducting transition is difficult to determine from the transport experiments as it does not coincide exactly with the bulk transition \cite{Rosuel2023, Sakai2023}. As shown in Ref.~\cite{Rosuel2023}, the anomaly in specific heat corresponding to the bulk transition to the hfSC phase appears as a hump and is extremely broad \cite{footnote2}. However, $H^\star$ is roughly independent of the sample quality \cite{Sakai2023}.  We also see in Fig.~\ref{comparisonpd}(a) that the value of the measured metamagnetic field $H_m$ varies from 33.8--34.75~T for the different samples. The lowest metamagnetic field has been found for the sample with the highest $\Tc$, \added{while in samples with a lower $\Tc$ a slightly higher $H_m$ has been observed.} \deleted[id=G]{A detailed analysis is difficult, as $H_m$ depends critically on the perfect field orientation along $b$ \mbox{\cite{Ran2019a, Helm2024}}.
A small misalignment from the $b$ axis results in an increase of $H_m (\theta) \propto 1/ \cos (\theta)$ in the $b-c$ plane and an even much stronger dependence in the $b-a$ plane   and the perfect aligment may not be guarantied in all the experiments. }
\added{The robustness of $H_m$ for different sample qualities has also been reported in Ref.~\cite{Wu2023}}

In Fig.~\ref{comparisonpd}(b) we plot $\Hc$ normalized by $\Tc$ as a function of $T/\Tc$. In this representation $\Hc (T/\Tc)$ in the lfSC phase scales on a single curve for samples with different $\Tc$, while in the high-field phase hfSC the critical field strong variations occurs depending on the samples. In general, in the clean limit, the orbital superconducting upper critical field at zero temperature depends on the Fermi velocities $v_{\rm F_\perp}$ perpendicular to the applied magnetic field $H$ and $\Tc$ as $\Hc^{\rm orb} \propto (\frac{\Tc}{v_{\rm F_\perp}})^2 $. 
However, in Fig.~\ref{comparisonpd}(b) we plot $\frac{\mu_0 \Hc}{\Tc}$ vs $\frac{T}{\Tc}$. This normalization is much better than that with 
a purely orbital limit $\frac{\Hc}{\Tc^2}$ vs $\frac{T}{\Tc}$ (see Fig.~\ref{Annex_comparison_PD} in the Appendix). As discussed in Ref.~\cite{Rosuel2023} the upper critical field for $H\parallel b$ is not described by a simple orbital limit. Surprisingly, as shown in Fig.~\ref{comparisonpd}, the low lfSC and hfSC phases show very different behavior for the different samples.  While the bulk nature of the hfSC phase has been clearly shown by specific heat, thermal expansion and ac susceptibility \cite{Rosuel2023, Kinjo2023PRB, Sakai2023}, the bulk transition itself is intrinsically very broad. The thermal expansion experiments clearly indicate that the vortex dynamics are different in the lfSC and hfSC phases \cite{Rosuel2023}. This is also supported by measurements of the critical current \cite{Tokiwa2023}, indicating that pinning in this phase depends strongly on sample quality and impurities. Here, we observe that $H^\star$ (Fig 6a), like $H_m$, appears weakly dependent on the sample purity, \cite{Thomas2021, Sakai2023} even though $\Tc$ changes very significantly between samples.

\section{$\mathbf{c}$ axis electrical transport under pressure}

\label{resistivity_pressure}

\subsection{Comparison of $\mathbf{c}$ axis and $\mathbf{a}$ axis resistivity in zero field}
\label{pressure_PD}

\begin{figure}[t]
	\includegraphics[width=0.8\linewidth]{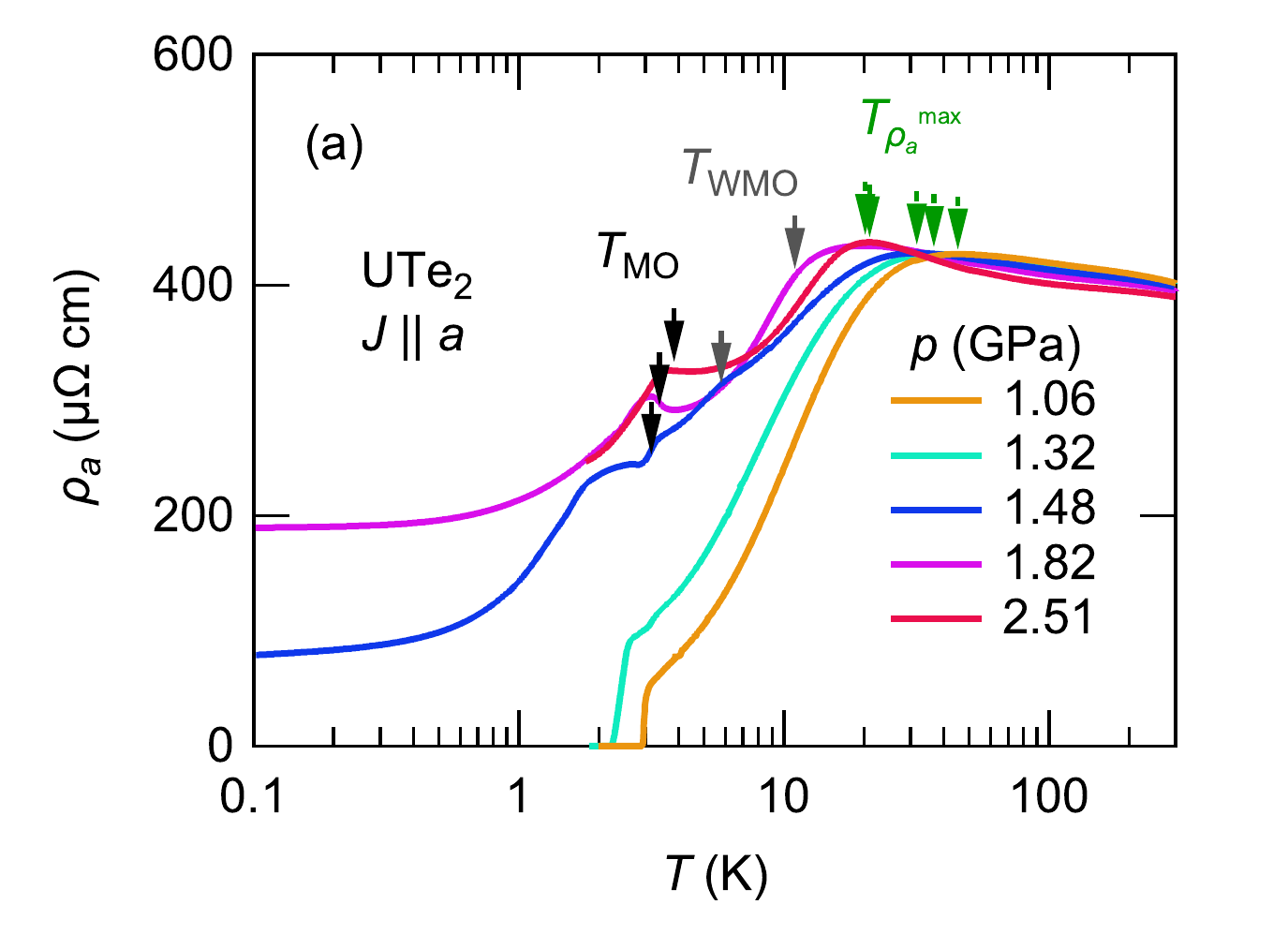}
	\includegraphics[width=0.8\linewidth]{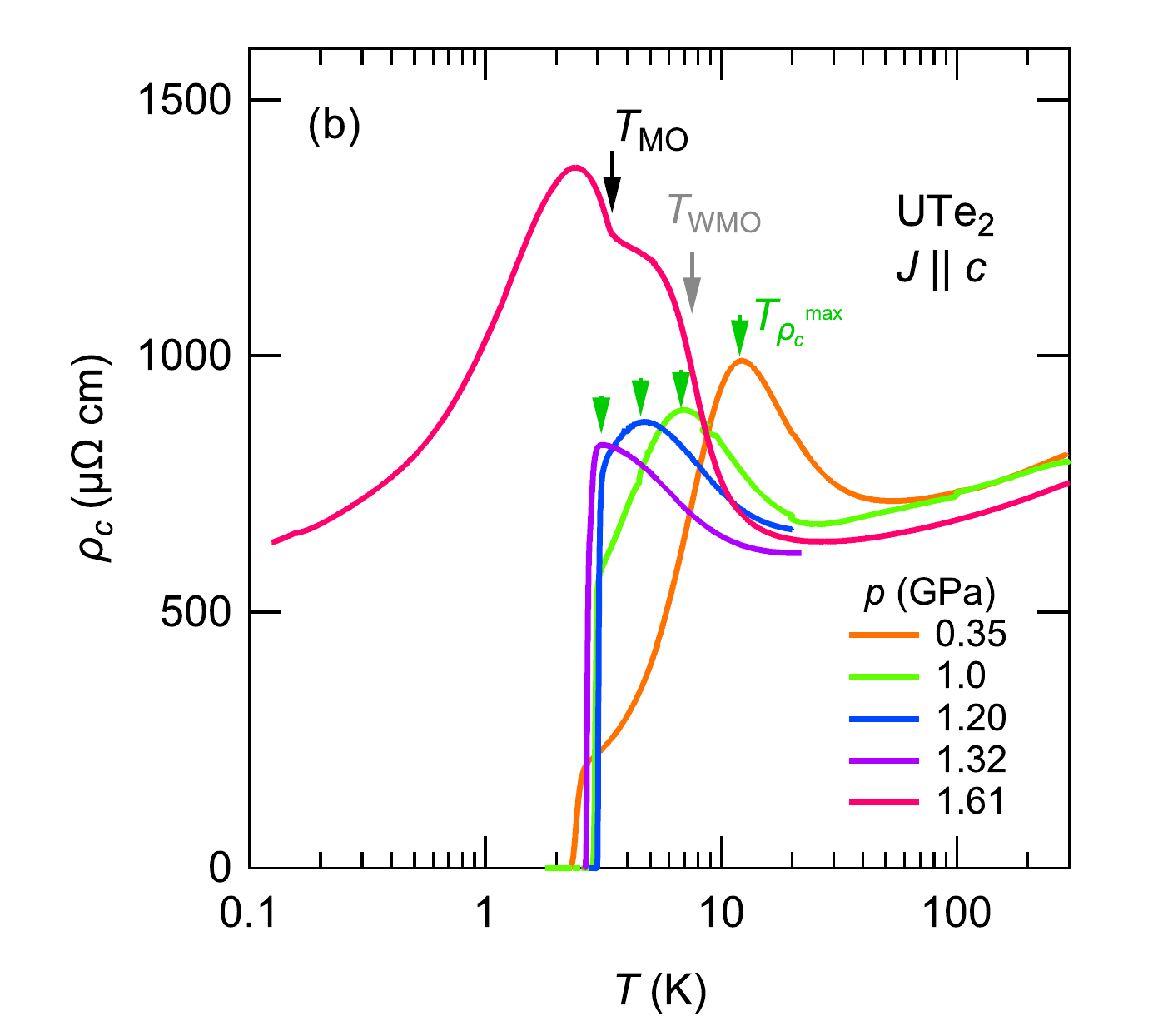}
	\caption{Temperature dependence of the resistivity for (a) $\rho_a$ (current $J || a$, data from the pressure experiment reported in Ref.~\cite{Knebel2020}) in UTe$_2$ at different pressures  on a logarithmic scale. The black and grey arrows indicate the ordering temperature $T_{\rm MO}$ and the crossover to a correlated regime $T_{\rm WMO}$. (b) Temperature dependence of $\rho_c$ (current $J || c$). The arrows indicate  $T_{\rm MO}$, $T_{\rm WMO}$, and $T^{\rm max}_{\rho_c}$.    }
	\label{Fig06}
\end{figure}

\begin{figure}[tb]
	\includegraphics[width=0.9\linewidth]{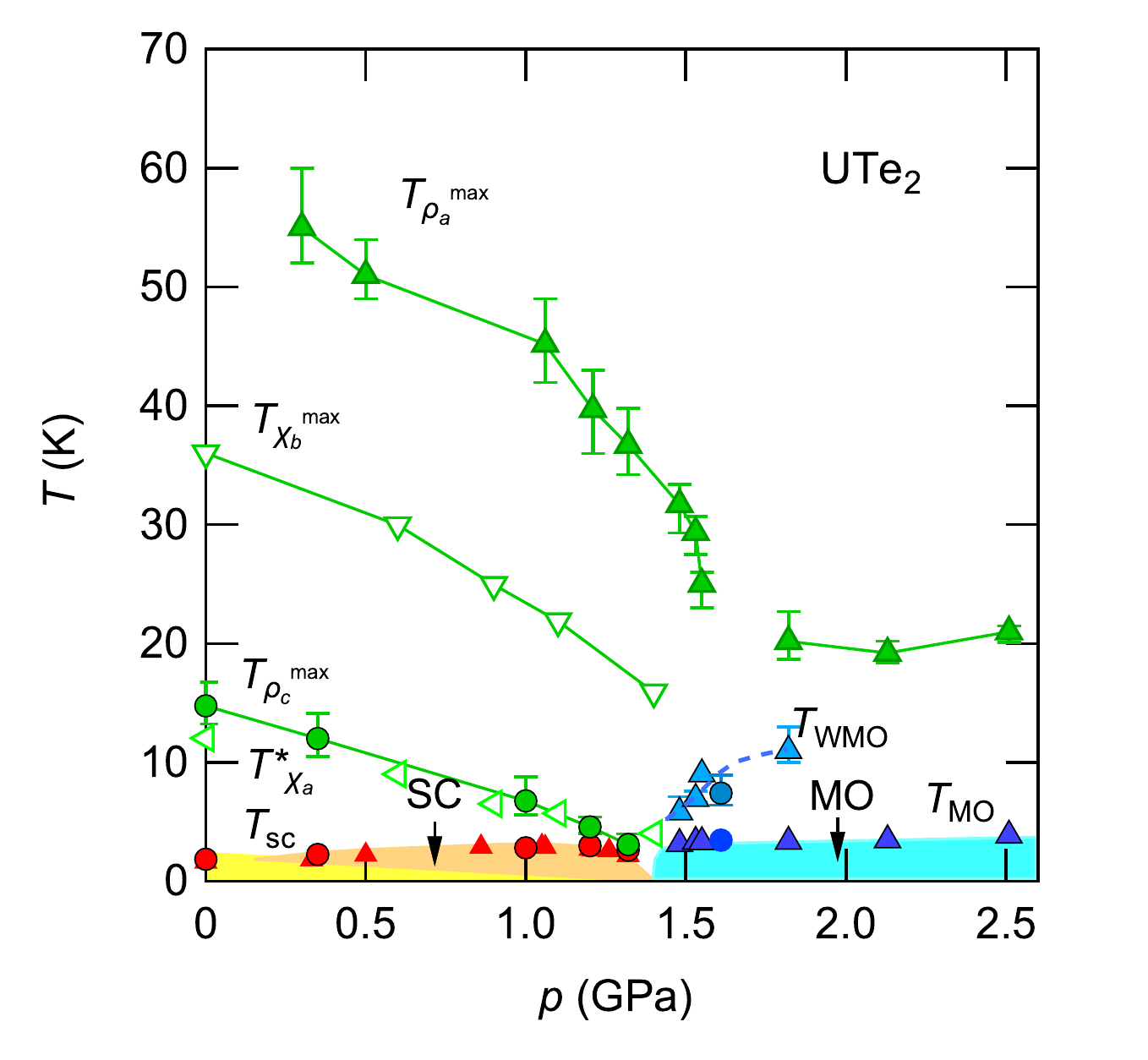}
	\caption{Pressure-temperature phase diagram of UTe$_2$ at zero magnetic field indicating the pressure dependence of $\Tmaxa$, $\Tchimax$ (taken from Ref.~\cite{Knebel2020}), $\Tmaxc$, $T^*_{\chi_a}$ (taken from Ref.~\cite{Li2021}), the superconducting transition temperature $T_{sc}$ (red symbols, triangles taken from Ref.~\cite{Knebel2020}), and the magnetic transition temperatures $T_{\rm MO}$ and $T_{\rm WMO}$. (Triangles-up and circles are from the $a$ and $c$ axis resistivity, respectively.)}
	\label{Fig07}
\end{figure}

Figure~\ref{Fig06} displays the temperature dependence of the resistivity with a current along the $a$ axis ($\rho_a$, upper panel) and along the $c$ axis ($\rho_c$, lower panel)  for different pressures at zero magnetic field. For both current directions, the general behavior at high temperature is similar to that at ambient pressure. $\rho_a (T)$ increases on cooling down to a temperature $\Tmaxa$ and decreases to lower temperatures.  The temperature of the broad maximum in $\rho_a$, $\Tmaxa$, decreases with increasing pressure, and is minimum near the critical pressure $p_c \approx 1.5$~GPa. At $p_c$ superconductivity is suppressed and the magnetically ordered ground state is formed (see Fig.~\ref{Fig07}). $\Tmaxa$   increases slightly above $p_c$.   The exact position of $p_c$ depends on the pressure conditions; $p_c \sim 1.7$~GPa has been reported in Refs.~\cite{Braithwaite2019} and \cite{Valiska2021} using anvil pressure cells. 

The temperature dependence of $\rho_a$ shows a kink for $p=1.48$~GPa near the temperature $T_{\rm WMO}\approx 6$~K and a second anomaly near $\TMO \approx3.5$~K in agreement with the previous report of two magnetic anomalies above $p_c$ \cite{Thomas2020, Aoki2021, Li2021}.  While at $\TMO$ long range magnetic order appears, $\TWMO$ has been identified in magnetization measurements under pressure as a crossover to a weakly magnetically ordered state \cite{Li2021}. With increasing pressure, $\TWMO$ shifts to higher temperature, gets less pronounced and cannot be resolved anymore above 2~GPa as it gets close to $\Tmaxa$. The lower anomaly at $\TMO$ is almost pressure independent up to 2.51~GPa, the highest pressure in this experiment (see Fig.~\ref{Fig07}). %
Very close to $p_c$ we observe at 1.48~GPa not only the magnetic anomaly, but also still the onset of a very broad superconducting anomaly. Similar coexistence of magnetism and superconductivity has been observed in  Ref.~\cite{Thomas2020} by ac calorimetry.  At the border of an antiferromagnetic instability, it is difficult to observe a magnetic transition inside the superconducting state ($\TMO < T_{sc}$), whereas superconductivity occurs often inhomogeneously inside the magnetically ordered state when $T_{sc} < \TMO$. A very well studied example for this competition of magnetic order and superconductivity is given by CeRhIn$_5$ \cite{Park2006, Knebel2006} with the inhomogeneous appearance of superconductivity below its critical pressure ($\Tc < T_N$ ), followed by the rapid suppression of magnetic order above $p_c$ when $\Tc > T_N$. 

As shown in Fig.~\ref{Fig06}(b), $\rho_c$ decreases with decreasing temperature for all pressures above 50~K. Similarly to the zero pressure data, at low temperatures a pronounced maximum $\Tmaxc$ occurs. This maximum shifts to lower temperatures with pressure up to 1.32~GPa. At this pressure, just below $p_c$, it almost coincides with the onset of superconductivity. The maximum of $T_{sc} \approx 3$~K is observed at 1.2~GPa for both samples. The normal-state resistivity changes drastically above the critical pressure $p_c$. At $ p = 1.61$~GPa, $\rho_c$ increases strongly below 10~K by a factor of two, shows a small plateau below 7~K and increases again below 3.45~K with a maximum at 2.4~K. The strong increase of the resistivity is attributed to a short range magnetic order at $T_{\rm WMO} \approx 7.5$~K and antiferromagnetic order below 3.5~K. A common feature between the $a$ and $c$ axes transport is the increase of the resistivity at the lower magnetic transition $\TMO$, which may indicate the opening of an electronic gap when entering in the magnetic state. The residual resistivity $\rho_0$ increases for both current directions strongly through $p_c$, and the anisotropy of the residual resistivity $\rho_{c0} / \rho_{a0}$ is still of the order of three in the magnetically ordered state. 
 
 In Fig.~\ref{Fig07} we summarize the pressure-temperature phase diagram of UTe$_2$ from the present resistivity measurements combined with previous studies \cite{Knebel2020, Li2021}. The characteristic temperatures determined from the temperature dependence of the resistivity $\Tmaxa$, $\Tmaxc$, the maximum of the magnetic susceptibility measured along the $b$ axis, $\Tchimax$, and $T^*_{\chi_a}$, which marks a broad shoulder in the susceptibility measured along the $a$ axis \cite{Li2021}, decrease up to the critical pressure $p_c \approx 1.5$~GPa.  $T^*_{\chi_a}$ follows roughly $\Tmaxc$.  Recently we have shown that $\Tmaxa$ scales with the maximum of the magnetic susceptibility $\Tchimax$ \cite{Valiska2021, Thebault2022} as a function of magnetic field and pressure, when a background contribution to the resistivity is subtracted. The background corresponds to the resistivity in the high-field regime above $H_m$, where the system is in a polarized state and magnetic fluctuations are strongly suppressed. 
 The magnetic interactions change drastically at the critical pressure and for $p> p_c$, the crystallographic $b$ axis becomes the axis of easy magnetization in the magnetic ordered state, while the $a$ axis is an intermediate axis \cite{Li2021, Kinjo2022PRB}.

\subsection{$\mathbf{c}$ axis transport under pressure and in magnetic field $\mathbf{H \parallel b}$}

\label{resistivity_P1-P5}

Next we focus on the $c$ axis resistivity $\rho_c$ for field applied along the $b$ axis at different pressures.  
 \begin{figure}[tb]
	\includegraphics[width=1\linewidth]{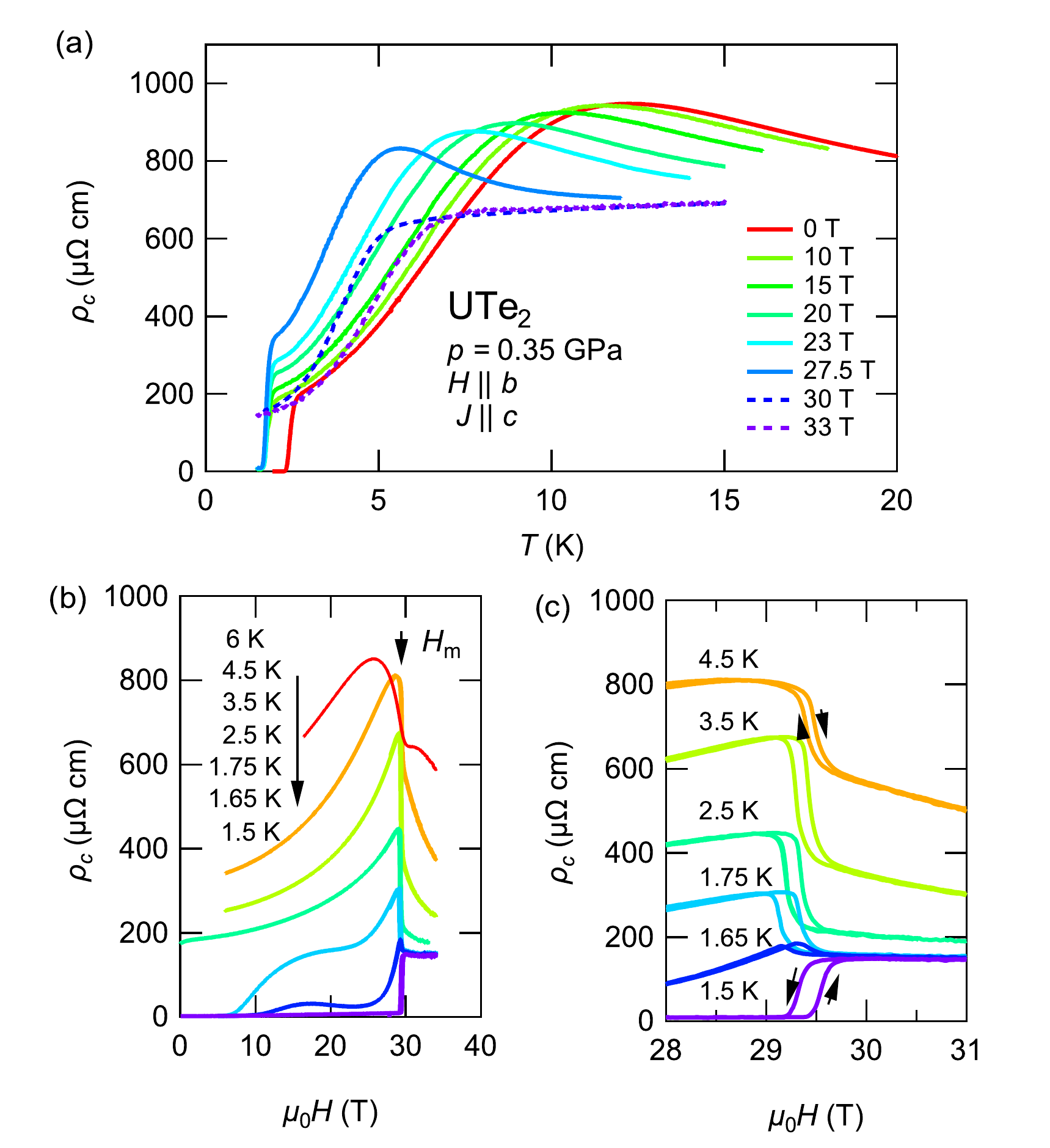}
	\caption{(a) Temperature dependence of the resistivity at 0.35~GPa for different magnetic fields. Solid (dashed) lines are for fields below (above) the metamagnetic transition. (b) Field dependence of the resistivity for different temperatures. (c) Zoom on the magnetoresistivity around the metamagnetic transition. }
	\label{Fig08}
\end{figure}
Figure~\ref{Fig08}(a) shows the temperature dependence of $\rho_c$ for different fields $H \parallel b$ at 0.35~GPa. Near $\Tmaxc \approx 12$~K the resistivity shows a maximum which is slightly lower than at ambient pressure. Under magnetic field, like at $p=0$ (see Fig.~\ref{Fig01}), the maximum shifts to lower temperatures and, at 27.5~T, we find $\Tmaxc \approx 5.5$~K. Above $H_m = 29.5$~T, at 30~T and for higher fields, the temperature dependence of $\rho_c$ changes significantly and instead of a maximum, a sharp drop of the resistivity is observed below 5~K. By further increasing the field, this anomaly is shifted to slightly higher temperatures. At low temperatures $\rho_c (T)$ has a $T^2$ temperature dependence in zero magnetic field up to 5~K, but with increasing field the $T^2$ range decreases significantly due to the low value of $\Tmaxc \approx 5$~K. Even above $H_m$ no clear $T^2$ is observed in the measured temperature range down to 1.5~K. 
The superconducting transition temperature decreases from  $\Tc = 2.28$~K at zero field down to $\Tc = 1.65$~K at 8~T and is almost field-independent in higher fields up to 27.5~T. Thus the field enhancement of superconductivity is less pronounced compared to ambient pressure, but a phase line between the lfSC and hfSC phases still exists \cite{Lin2020}.  

 In Fig.~\ref{Fig08}(b) we show the field dependence $\rho_c (H)$ up to 35~T for different temperatures at $p = 0.35$~GPa. At the lowest temperature, superconductivity survives up to $H_m = 29.5$~T. A large jump to the normal state resistivity occurs at $H_m$. $\rho_c (H)$ shows a hysteresis between the up and down sweeps of the field [see Fig.~\ref{Fig08}(c)] indicating the first order nature of the transition at $H_m$. At 1.65~K,   the signature of the first order metamagnetic transition is very weak.  At higher temperatures, we observe a marked hysteresis at the metamagnetic transition and the magnetoresistance $\rho_c (H)$ decreases strongly. At 6~K no hysteresis is observed anymore, indicating that the critical point of the first order transition is located between 4.5--6~K, thus lower than at ambient pressure.  

 \begin{figure}[tb]
	\includegraphics[width=1\linewidth]{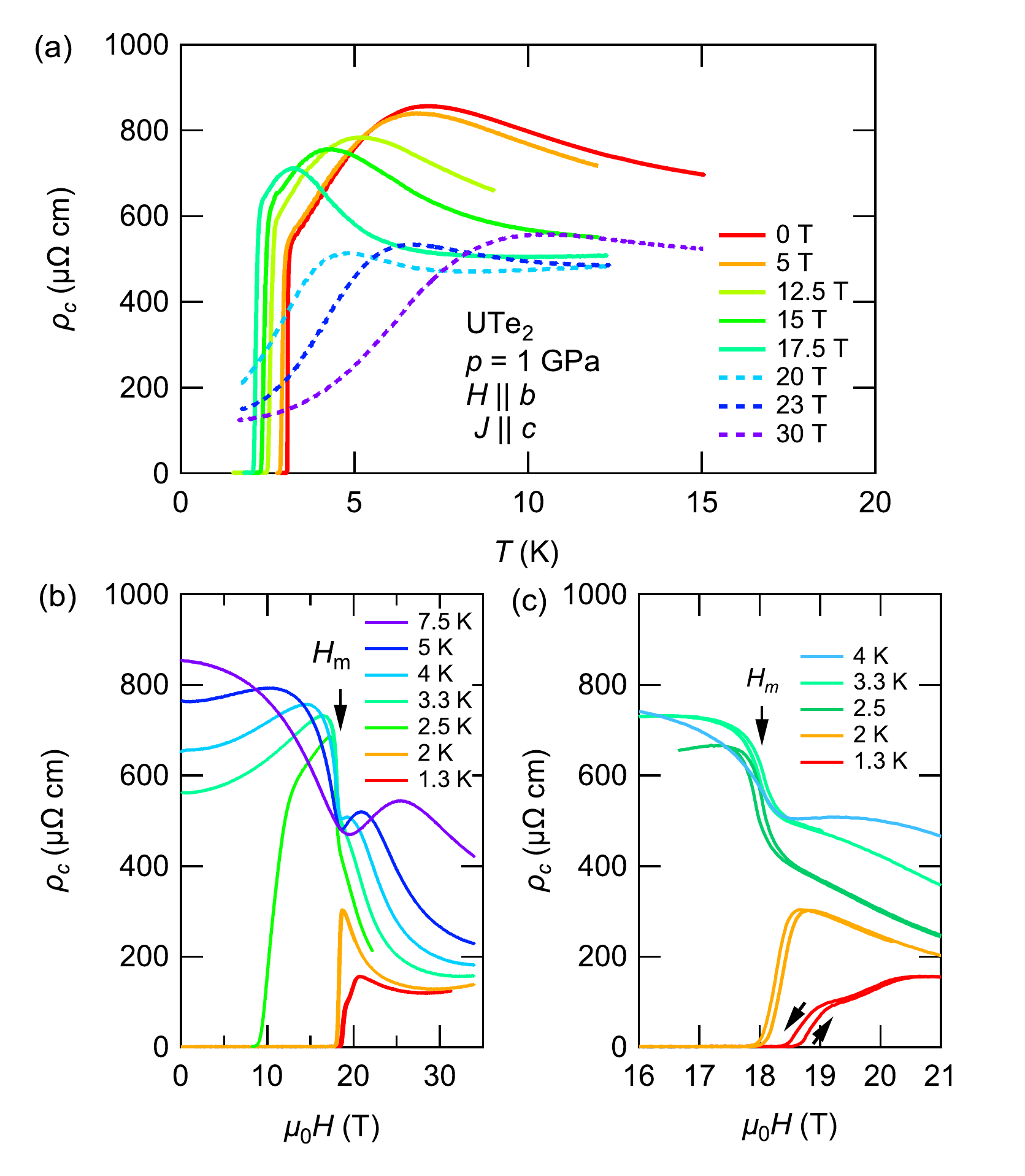}
	\caption{(a) Temperature dependence of the resistivity at 1~GPa for different magnetic fields. Solid (dashed) lines are for fields below (above) the metamagnetic transition. (b) Field dependence of the resisitivity for different temperatures. (c) Zoom on the magnetoresistivity around the metamagnetic transition. }
	\label{Fig09}
\end{figure}

\begin{figure}[tb]
	\includegraphics[width=1\linewidth]{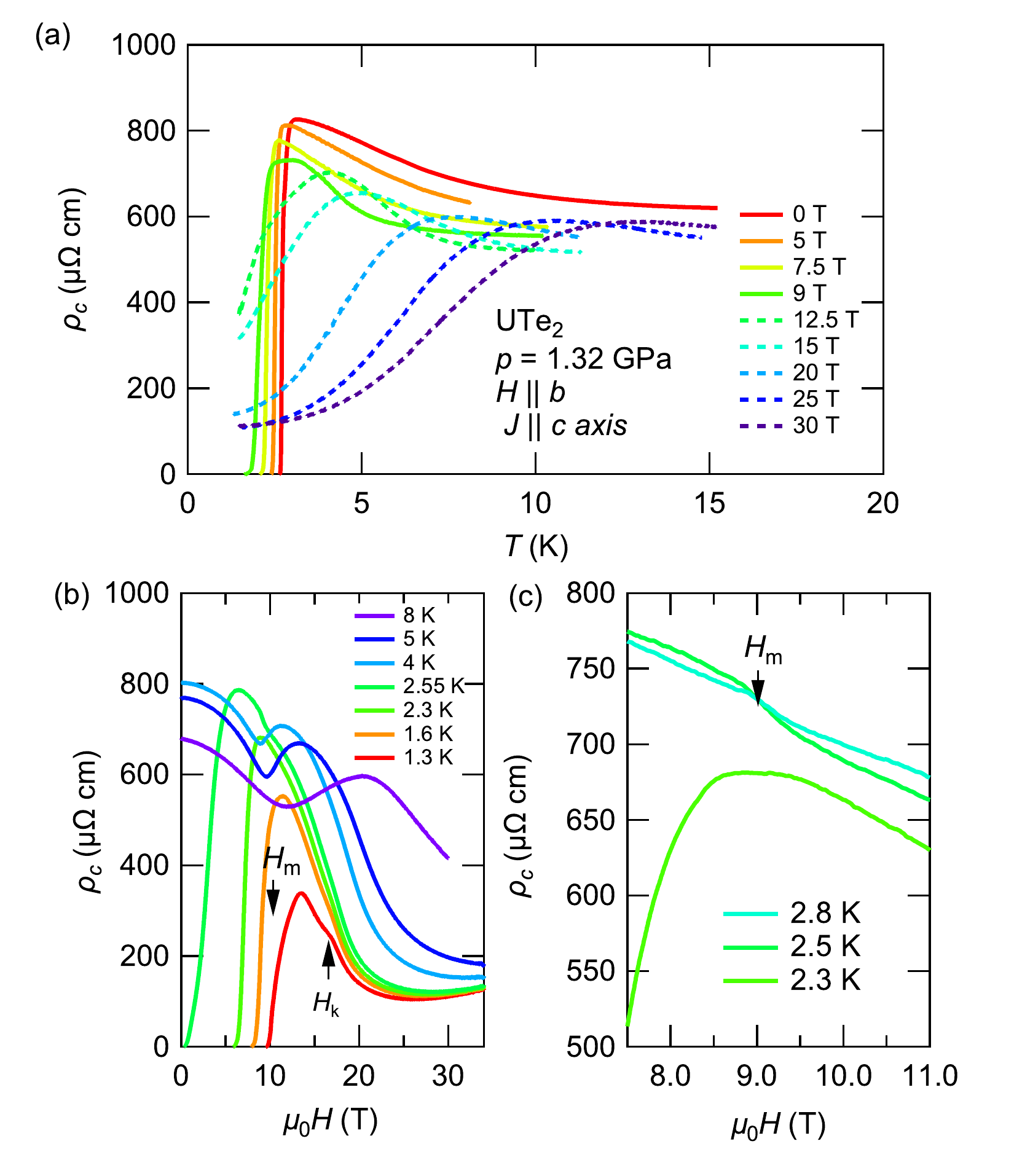}
	\caption{(a) Temperature dependence of the resistivity at 1.32~GPa for different magnetic fields. Solid (dashed) lines are for fields below (above) the metamagnetic transition. (b) Field dependence of the resistivity for different temperatures. (c) Zoom on the magnetoresistivity around the metamagnetic transition. }
	\label{Fig10}
\end{figure}

Figures \ref{Fig09} and \ref{Fig10} show the resistivity for higher pressures (1 GPa and 1.32 GPa respectively) approaching the critical pressure. 
From the magnetoresistance at 1~GPa we conclude that the metamagnetic transition field is further reduced down to $\mu_0 H_m = 18.5$~T [see Fig.\ref{Fig09} (b)]. The temperature dependence of the resistivity has changed slightly compared to lower pressures, and only the characteristic temperatures are lower. The maximum in $\rho(T)$ decreases from $\Tmaxc= 7.5$~K at $H = 0$  to 3.24~K at $\mu_0 H =17.5$~T. For fields above $H_m$ the absolute value of the resistivity has decreased and a shallow maximum in the temperature dependence appears at 4.8~K at 20~T. This maximum shifts to higher temperatures with increasing magnetic field. The superconducting transition is monotonously suppressed from 3.05~K at $H=0$ to  2.07~K at 17.5~T. Due to the high superconducting transition temperature and to the low $\Tmaxc$, no $T^2$ temperature dependence is observed in the normal state above the superconducting transition. A Fermi-liquid $T^2$ dependence is only recovered above the metamagnetic transition in the polarized phase for fields above $H_m$, but the maximum temperature is about 3.5~K, thus the $T^2$ regime is very narrow.
We observe zero resistivity up to $H_m$ at 1.3 and 2~K [see Fig.~\ref{Fig09}(b,c)]. We do not find any signature of a reinforcement of superconductivity, but the upper critical field is monotonously suppressed.

In the normal state a maximum develops in the magnetoresistance at temperatures higher than the critical point, which is near 3.5~K and 17.5~T at 1~GPa. The maximum shifts to higher temperatures with increasing magnetic fields. 

Figure~\ref{Fig10} presents the temperature and field dependence of $\rho_c$ close to the critical pressure at 1.32~GPa. At zero magnetic field the resistivity increases with decreasing temperature, superconductivity sets in near 3~K and zero resistivity is attained at 2.65~K. The shallow maximum near 3.05~K at zero field may already be a signature of the onset of superconductivity. The characteristic temperature $\Tmaxc$ seems to be lower or at most of the same order than $T_{sc}$. Under magnetic field up to 9~T a shallow maximum occurs just above the onset of superconductivity.  Above  $H \approx 9$~T, close to $H_m$, a broad maximum in $\rho_c (T)$ occurs and is shifted to higher temperatures for higher magnetic fields. At low temperatures a Fermi-liquid regime occurs. 

\begin{figure}[tb]
	\includegraphics[width=1\linewidth]{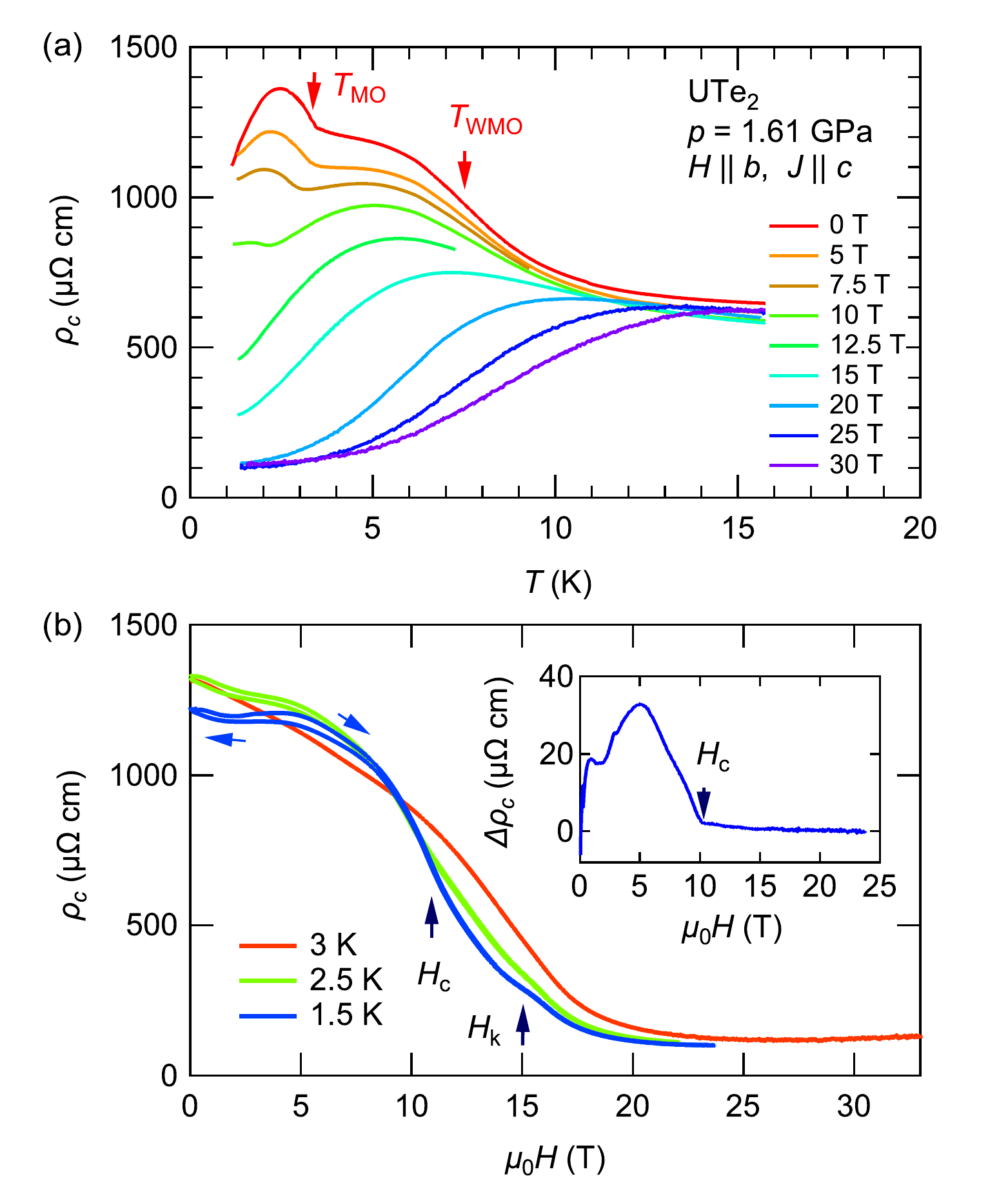}
	\caption{(a) Temperature dependence of the resistivity at 1.61~GPa for different magnetic fields. $T_{\rm MO}$ indicates the magnetic ordering temperature and $T_{\rm WMO}$ the crossover to a short-range correlated regime at zero field, respectively. (b) Field dependence  of the resistivity for different temperatures measured at the LNCMI. Blue arrows indicate the direction of the field sweep. $H_c$ and $H_k$ mark the critical field of the disappearance of the hysteresis and a well defined kink in the magnetoresistivity at 1.5~K. The inset in (b) shows the difference $\Delta \rho_c$ between field up and field down sweeps at 1.5~K, with the opening of the hysteresis at $H_c = 10.3$~T. }  
	\label{Fig11}
\end{figure}

\begin{figure}[htb]
	\includegraphics[width=1\linewidth]{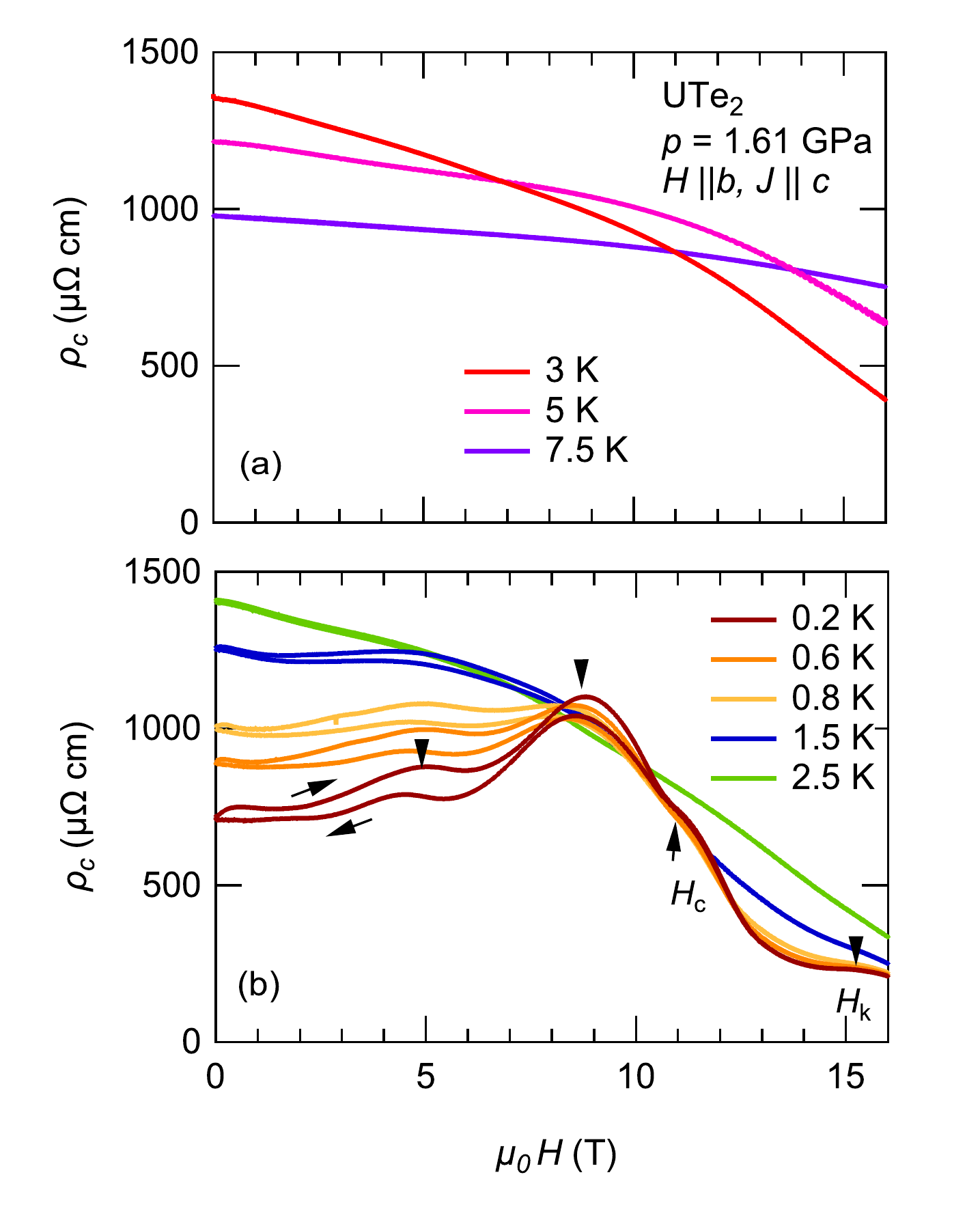}
	\caption{Field dependence of the $c$ axis resistivity at 1.61~GPa for different temperatures above (a) and below $T_M$ (b). A strong hysteresis occurs in the magnetic state. $H_c$ indicates the critical field of the magnetic order. $H_k$ indicates a kink in the resistivity (see also Fig.~\ref{Fig11} above).  }
	\label{Fig12}
\end{figure}

\begin{figure*}[htb]
	\includegraphics[width=1\linewidth]{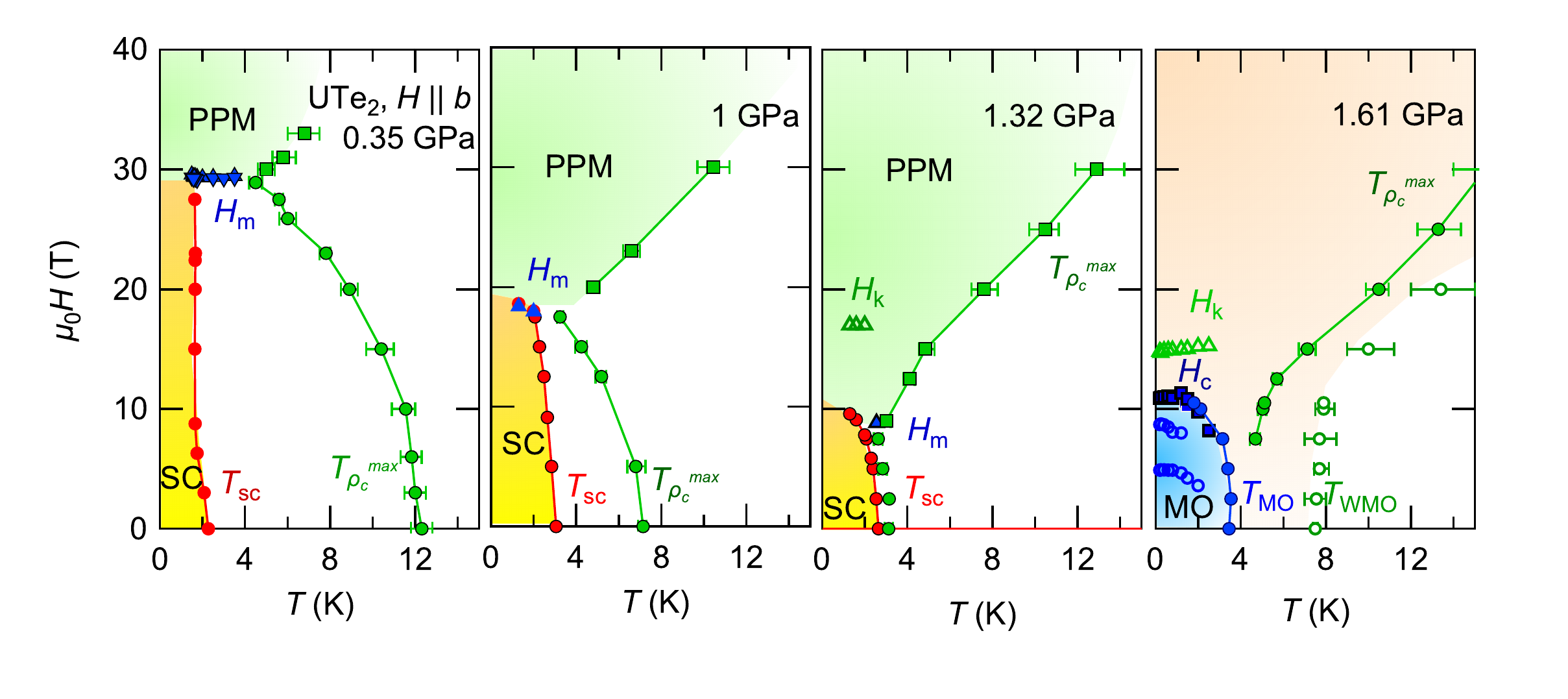}
	\caption{Magnetic and superconducting phase diagram of UTe$_2$ at 0.35, 1.1, 1.32, and 1.61~GPa  for field applied along the $b$ axis defined from the resistivity measurements with $J \parallel c$. Full green circles (squares) indicate the maximum of the resistivity $\Tmaxc$ as function of temperature below (above) the metamagnetic transition. The superconducting transition at $T_{sc} (H)$ (full red circles) is defined by zero resistivity. 
		The metamagnetic transition at $H_m$ is marked by blue triangles for pressures below $p_c \approx 1.45$~GPa. SC marks the superconducting phase, PPM the polarized paramagnetic phase above $H_m$. At 1.61~GPa long range magnetic order (MO) occurs below $\TMO$ up to $H_c$. $\TWMO$ (open circles) is defined by the inflection point of the increase of the $c$ axis resistivity and shows a similar field dependence as the maximum of the resistivity $\Tmaxc$. No metamagnetism occurs. 
		Open blue circles  Open blue circles (determined from extrema of the magnetoresistivity) correspond probably moment reorientations in the MO state. $H_c$ indicates the critical field for the magnetically ordered state. $H_k$ indicates the kink in the magnetoresistivity in the polarized high field regime.  
	}
	\label{Fig_phase_diagrams}
\end{figure*}

The field dependence $\rho_c (H)$ is shown in Fig.~\ref{Fig10}(b) for different temperatures. Panel (c) displays the field range around the metamagnetic transition in an enlarged scale. We first discuss $\rho (H)$ at the lowest temperature $T = 1.3$~K [see Fig.~\ref{Fig10}(b)]. Zero resistivity is observed up to $\approx 9.6$~T in the field sweep with increasing field. For higher fields the resistivity increases sharply and has a change of slope as a function of field near 10.15~T. A clear signature of the metamagnetic transition is missing. 
By further increasing the field, $\rho_c (H)$ at 1.3~K has a maximum at 13.3~T, which marks the onset of superconductivity near $\Hc$ in the field sweep. A well defined kink in $\rho_c (H)$ occurs at  $H_k = 17.1$~T in the normal state. For even higher fields the magnetoresistivity has a minimum around 26~T.  With increasing temperature these features get broadened and the kink cannot be followed above 2~K. Figure \ref{Fig10}(c) displays the anomaly of the metamagnetic transition on an enlarged scale. While at 2.3~K almost no anomaly at the metamagnetic transition occurs due to the closeness to the onset of superconductivity (at 8.7~T), at 2.5 and 2.8~K only a tiny anomaly indicates the metamagnetic transition; no jump or marked decrease in $\rho_c (H)$ occurs anymore. Therefore the critical point of the first order transition can no more be resolved due to superconductivity.  For temperatures above 2.8~K no signature of $H_m$ is detected. At 4~K the magnetoresistivity shows a minimum at 9~T and a maximum at 11.4~T. This maximum shifts to higher fields with increasing temperature, similarly to the temperature $\Tmaxc$ of the maximum in $\rho_c (T)$ at fixed field. 

Finally, in Figs.~\ref{Fig11} and \ref{Fig12} we show the temperature and field dependencies of $c$ axis resistivity at $p = 1.61$~GPa.  On cooling the resistivity shows a smooth increase at zero field below 10~K with an inflection point at $T_{\rm WMO} \approx 7.5$~K, and a sharp increase at $T_{\rm MO}= 3.45$~K. Under field, $T_{\rm WMO}$ increases slightly in temperature up to 10~T, and much faster for higher fields when it gets less pronounced, as the absolute value of the resistivity decreases (see Fig.~\ref{Annex_dif_P5} in the Appendix which shows $d\rho / dT$ for different magnetic fields). On the contrary $T_{\rm MO}$ decreases with increasing field, and its signature stays sharp under field up to 10~K. 

A strongly negative magnetoresistance up to fields of 30~T is visible in Fig.~\ref{Fig11}(b). It decreases by more than a factor 10 from zero field to 20~T indicating the suppression of the magnetic scattering related to the magnetic correlations established below the characteristic temperature $T_{\rm WMO}$. This strong magnetoresistance at low temperatures contrasts with the rather low variation of the magnetoresistance by only 2\% at 15~K. For temperatures above  $T_{\rm MO}$ the magnetoresistance does not show any hysteresis and decreases monotonously below 20~T. On the contrary, below $T_{\rm MO}$ the magnetoresistance shows a significant hysteresis for field up and field down sweeps. We define the field $H_c$, where the hysteresis disappears  as the critical field of the magnetically ordered state, see inset in Fig.~\ref{Fig11}. $H_c$ determined here from the field sweeps coincides with the temperature $T_{\rm MO}$ determined from the temperature sweeps. Inside the magnetically ordered state several additional anomalies occur. In Figs.~\ref{Fig11}(b) and \ref{Fig12}(b) maxima of the magnetoresistance are found in fields $H < H_c$. These findings are similar to those reported in Refs.~\cite{Lin2020, Valiska2021}. The observation of clear anomalies in the magnetic ordered state is presumably an indication for re-orientations of the magnetic moments inside the magnetically ordered state. Of course, thermodynamic measurements are needed to determine the phase lines inside the ordered state. 
Very recently, an incommensurate antiferromagnetic order has been observed by neutron diffraction experiments \cite{Knafo2023}.

 For fields $H > H_c$ we observe at low temperatures an additional kink at $H_k$ in the magnetoresistance. A similar feature had been already observed at lower pressures in the polarized state at $p = 1.32$ GPa. The signature of $H_k$ disappears for temperatures above 3~K. A comparable kink in the magnetoresistance has been observed for an angle of 25~deg from the $b$ toward the $c$ axis at about 21~T \cite{Ran2021}. The origin of this anomaly is still an open question.

\subsection{Discussion of the pressure dependence}

\label{discussion}

\begin{figure}[htb]
	\includegraphics[width=1\linewidth]{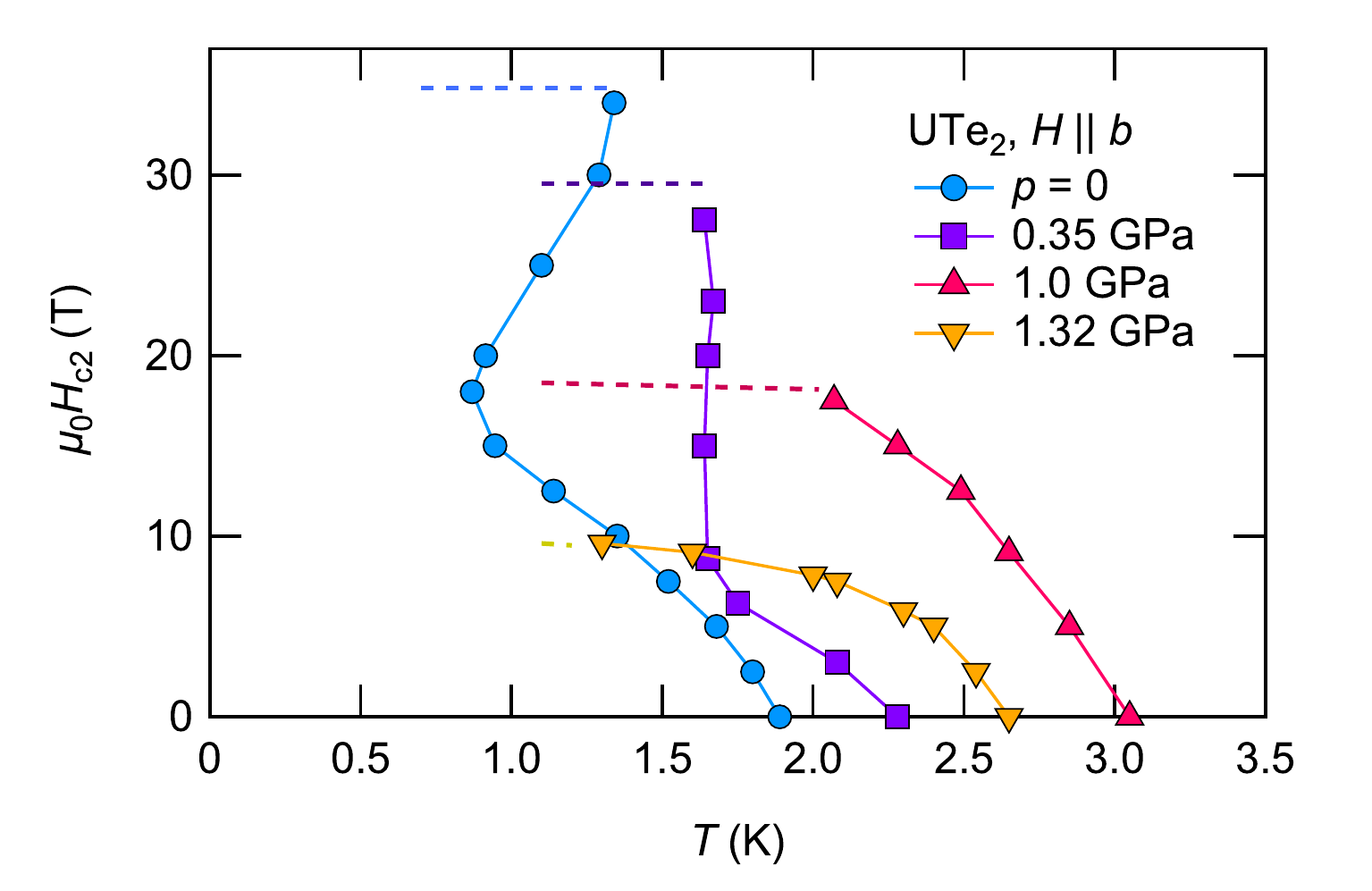}
	\caption{Superconducting upper critical field $\Hc$ for different pressures. Dashed lines indicate the metamagnetic field for each pressure, which is the upper field limit of superconductivity. }
	\label{Hc2_pressure}
\end{figure}

\begin{figure}[htb]
	\includegraphics[width=1\linewidth]{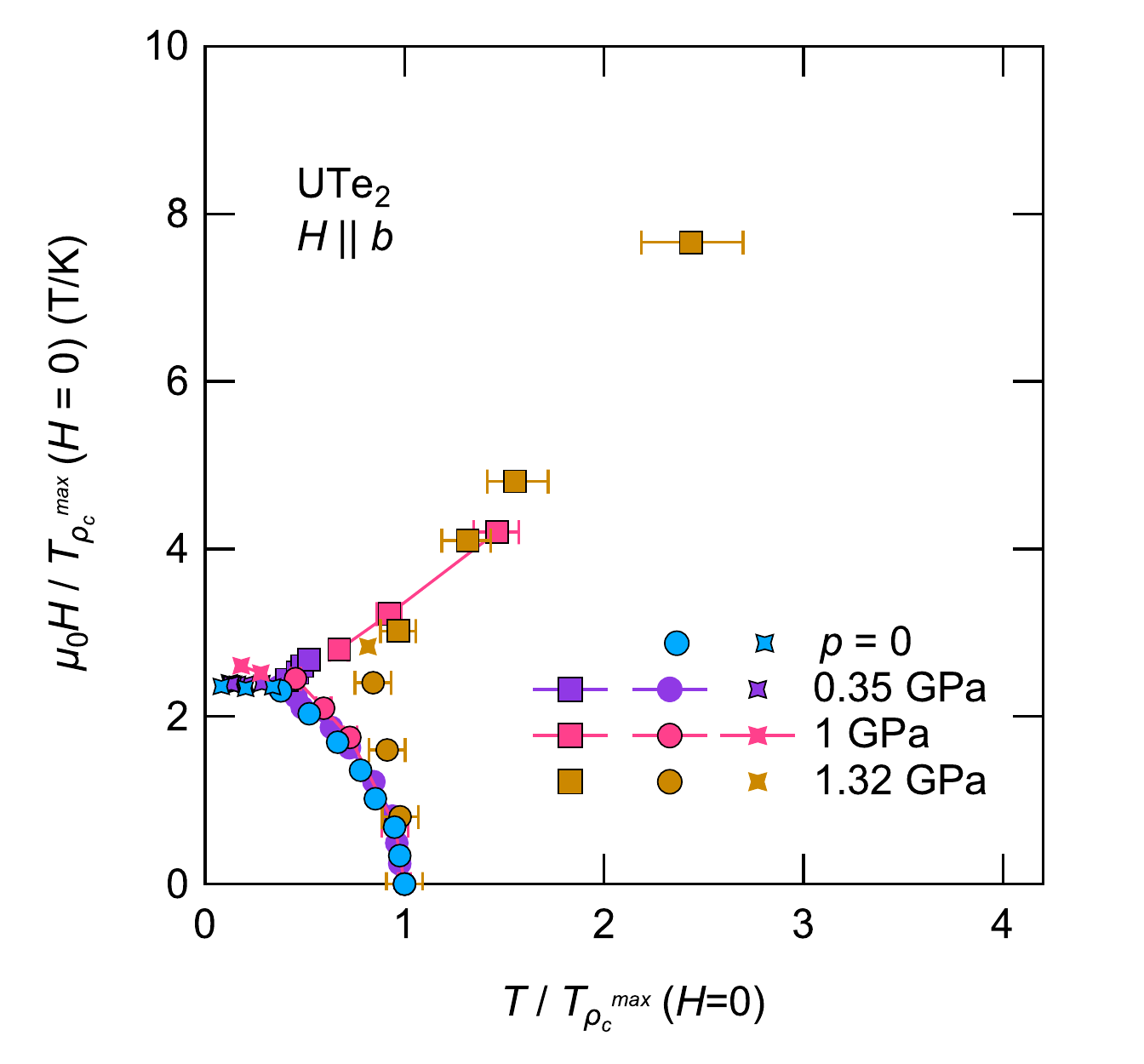}
	\caption{Magnetic phase diagram of UTe$_2$ for field along the $b$ axis normalized to the temperature $\Tmaxc (H=0)$ at zero magnetic field as a function of the normalized temperature. Circles (squares) give $\Tmaxc$ below (above) the metamagnetic transition, stars indicate $H_m (T)$ for the different pressures.   }
	\label{Tmax_scaled}
\end{figure}

Figure~\ref{Fig_phase_diagrams} summarizes the $H-T$ phase diagrams at different pressures determined from the resistivity measurements with current $J \parallel c$ and field $H \parallel b$. For all pressures below $p_c$, $\Tmaxc$ is  connected to the metamagnetic transition. Superconductivity is observed up to $H_m$, which decreases with pressure. For fields higher than $H_m$ a crossover to a polarized regime occurs as a function of temperature. Close to the critical pressure an additional kink occurs in the polarized state at low temperatures in $\rho_c (H)$ at $H_k$, which persists also above $p_c$.  Above the critical pressure, at 1.61~GPa, two magnetic anomalies occur: the lower anomaly $\TMO$ is a transition to a long range magnetically ordered state. Its antiferromagnetic nature was suggested by the different phase lines inside the ordered state and has been recently shown by neutron diffraction under high pressure \cite{Knafo2023}. On the contrary, the presence of a true phase transition at $\TWMO$  is less clear, and not only broad features occurs in the resistivity, but also in magnetization \cite{Li2021} or specific heat measurements \cite{Thomas2020}.  The field dependence of $\TWMO (H)$ resembles that of a ferromagnet under field applied along the easy magnetization axis, as the transition at $\TWMO$ smears out under a magnetic field and shifts to higher temperatures in a similar way as the maximum $\Tmaxc$. However, the microscopic origin of $\TWMO$ has not been identified today, crystal field effects may also be not negligible. We recall that the magnetic anisotropy changes at $p_c$ with $b$ being the easy and $c$ the hard magnetization axis \cite{Li2021}. The coexistence and competition of antiferromagnetic and ferromagnetic fluctuations has been suggested from NMR  measurements \cite{Ambika2022}.

 The pressure evolution of the superconducting phase diagram is similar to those previously discussed in Ref.~\cite{Knebel2020} and shown in more detail in Fig.~\ref{Hc2_pressure}. The superconducting transition temperature increases with applied pressure and superconductivity is stable under magnetic field up to the metamagnetic transition, but the temperature dependence of the upper critical field $\Hc (T)$ changes significantly under pressure. At low pressure (see at 0.35~GPa) a reinforcement of superconductivity \added{under magnetic field} can still be observed. The critical field $H^\star$ of the upturn of  $\Hc (T)$ is  already strongly reduced from about $17$~T at zero pressure to around $8$~T at 0.35~GPa. The pressure 0.35~GPa is very close to the pressure, where ac calorimetry  at zero field indicates two different superconducting phases \cite{Braithwaite2019, Aoki2020, Thomas2020}. \added{Previous reports of the upper critical field for $H \parallel b$ \cite{Knebel2020, Lin2020} in that pressure range  show at low field a concave curvature indicating the increase of pairing strength under magnetic field. As initially $\Tc$ decreases with increasing pressure, the data at 0.35~GPa suggest that for this pressure already two superconducting phases should exist, since $\Tc$ is higher than at zero pressure. Thus, on cooling at zero magnetic field we should enter first in a high-temperature superconducting phase. Of course,} 
 with resistivity measurements, it is not possible to discern different superconducting phases and detect phase lines inside a superconducting phase. The critical temperature $\Tc$ determined by resistivity measurements is always that of the phase with the higher $\Tc$. \added{In Refs.~\cite{Lin2020, Kinjo2023} it has been proposed that the lfSC is suppressed with pressure and embedded in the high-temperature superconducting phase. The high-temperature superconducting phase is proposed to be the pressure continuation the hfSC.}
  \deleted[id=G]{Under pressure the lfSC is suppressed and embedded  in the high-field superconducting phase hfSC, which  survives at $\Tc$ up to $p_c$, as proposed in Ref.~\cite{Lin2020}. }
 More detailed studies by thermodynamic probes of the phase line between the lfSC and hfSC phases for $H\parallel b$  are needed in future. \added{They should be performed in the vicinity of the pressure where the high-temperature superconducting phase appears and under magnetic field applied along $b$.  }

 Obviously, the temperature dependence of $\Hc$ changes on approaching $p_c$. At $p=1$~GPa we do not observe any signature of a field enhancement of superconductivity up to $H_m$ \added{in agreement with previous studies \cite{Knebel2020, Lin2020}}. \deleted[id=G]{It has been proposed that lfSC occurs inside the hfSC phase \mbox{\cite{Lin2020, Kinjo2023}}, and that the hfSC corresponds to the high-temperature superconducting phase under pressure.} The strong curvature of $\Hc (T)$ might indicate the increase of the paramagnetic limitation near $p_c$ for $H \parallel b$ \cite{Knebel2020}. The question of the pairing symmetry of the superconducting state near $p_c$ is still open. \added{On one side, the NMR Knight shift stays constant on cooling through $\Tc$ favoring an odd parity pairing. On the other side, if the high-temperature superconducting phase corresponds to the pressure evolution of the hfSC state, it could also be a singlet state.}  
 In Ref.~\cite{Rosuel2023} it has been proposed that the hfSC may be spin singlet and driven by critical fluctuations on approaching the metamagnetic transition. Under pressure, the hfSC state may survives up to $p_c$.  It has even been predicted \cite{Ishizuka2021} that antiferromagnetic fluctuations would be reinforced under pressure favoring a spin-singlet state for the high-temperature superconducting phase above 0.3~GPa. 
  As the magnetically ordered state MO is antiferromagnetic, antiferromagnetic fluctuations may drive the superconductivity close to the critical pressure. \added{However, even at ambient pressure and low field the pairing mechanism in UTe$_2$ is still not settled.}
 
Next we will concentrate on the normal state properties. As shown in Fig.~\ref{Fig_phase_diagrams} the temperature $\Tmaxc$ of the maximum in the $c$ axis resistivity  is decreasing with applied pressure. For all pressures below $p_c$ the field dependence of $\Tmaxc$ connects to the metamagnetic transition $H_m$, where $\Tmaxc$ has its lowest value. For fields above $H_m$, $\Tmaxc$ marks the crossover to a polarized paramagnetic regime. If we identify the minimum value of $\Tmaxc$ as the critical point of the first order metamagnetic transition, we determine that this critical point decreases almost linearly with pressure from about (5.6~K, 34~T) at zero pressure to (2.55~K, 8.85~T) at 1.32 GPa. At 1.32~GPa the signature of the metamagnetic transition is almost lost and only a tiny anomaly is visible.

\begin{figure}[t]
	\includegraphics[width=1\linewidth]{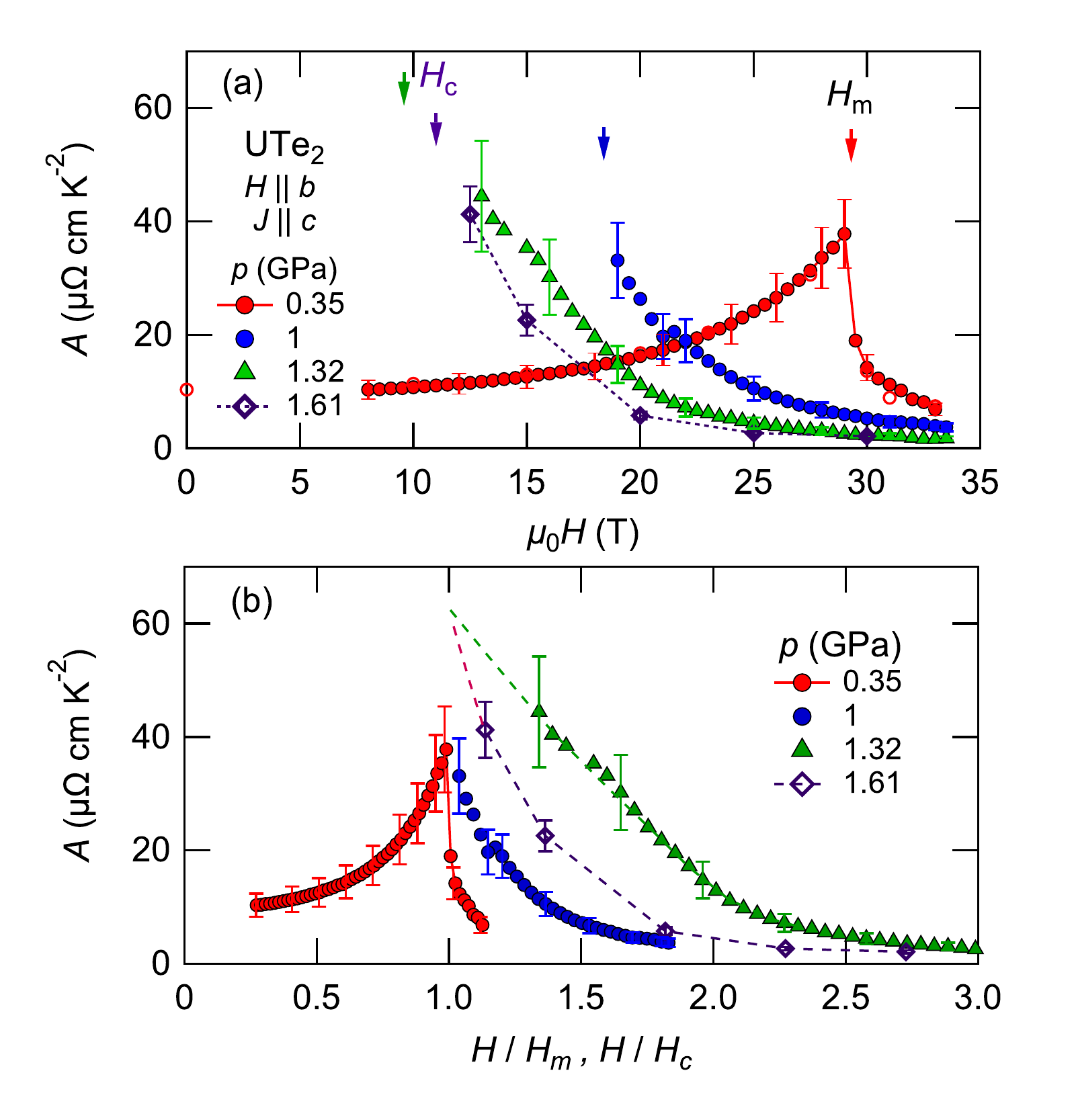}
	\caption{(a) $A$ coefficient as a function of magnetic field. Solid arrows indicate the position of $H_m$ for the different pressures, the dashed arrow the critical field of the magnetic order. 
	Due to the closeness of $\Tmaxc$ and $\Tc$, $A$ can be only determined at 1 and 1.32 GPa for $H > H_m$, where superconductivity is suppressed. 
	Above the critical pressure at 1.61~GPa a Fermi liquid is only observed for fields above $H_c$. Dashed lines are guides to the eye. 
     (b) $A$ as a function of the normalized field, $H_m$ for $p < p_c$ and $H_c$ for $p > p_c$. Lines are guides to the eye. }
	\label{A_coefficient_pressure}
\end{figure}

In Fig.~\ref{Tmax_scaled} we plot the phase diagram of the normal state scaled in field and temperature by the temperature $\Tmaxc (H=0)$. Importantly,  the phase diagrams of the different pressures below $p_c$ scale almost perfectly. Only for $p = 1.32$~GPa the scaling is less good, but this may be due to the difficulty to determine $\Tmaxc$ at low field correctly due to the high superconducting temperature, $\Tc \approx \Tmaxc$. Our thermal expansion measurements at zero pressure clearly show the correspondence  $T^* \approx T^\star_\alpha \approx \Tmaxc \approx 14.5$~K. This characteristic temperature scale connects to the metamagnetic transition at $H_m$. The microscopic origin of $T^\star$ is clearly related to the interplay of magnetic fluctuations and the formation of a coherent heavy-fermion state. This has been shown by different NMR and also inelastic neutron scattering experiments at ambient pressure. Inelastic neutron scattering evidenced the development of antiferromagnetic fluctuations at wave vector $\mathbf{k_1} = (0, 0.57, 0)$  below 60~K which saturate below 15~K \cite{Duan2020, Knafo2021, Butch2022}. The magnetic fluctuations in UTe$_2$ depend strongly on the particular ladder like structure of the U atoms along the $a$ axis with the rung along $c$ axis. Two dimensional antiferromagnetic fluctuations originating from  magnetic ladders coupled along $b$ were captured by inelastic neutron scattering and have a characteristic energy scale of 3-4~meV. The temperature dependence of these fluctuations is compatible with that from the NMR relaxation rate $1/T_{1}T$ \cite{Tokunaga2019, Tokunaga2022}. Under pressure,  $1/T_1T$ measured on the different Te-sites, also show a constant behavior up to $T^\star$ \cite{Ambika2022, Kinjo2022PRB} and  $T^\star$ from the NMR scales with $\Tmaxc$ determined from the resistivity. This seems to be the dominant energy scale which determines the pressure and field dependence of the phase diagram. 

We emphasize that electronic correlations increase on approaching the critical pressure. The field dependence of the $A$ coefficient from our measurements with a current along the $c$ axis is shown in Fig.~\ref{A_coefficient_pressure}(a) as a function of the field and in (b) as a function of the normalized field $H/H_m$ ($p< p_c$) or $H/H_c$ ($p> p_c$). (Figs.~\ref{Annex_rho_T2_pressure} and \ref{Annex_rho_T2_P5} in the Appendix show the experimental data and the corresponding fits.)  At 0.35~GPa $A(H)$ has still a similar field dependence as at ambient pressure and $A$ shows a steplike decrease just above the metamagnetic transition. For higher pressures, due to the lower $\Tmaxc$, the onset of superconductivity prevents the determination of $A$ below the metamagnetic transition in $\rho_c$. Thus $A(H)$ has been only determined in the field range above $H_c$. It shows a strong monotonous decrease. The enhancement of the $A$ coefficient at $H_m$ is the largest close to the critical pressure and comparable to $A$ on approaching $H_c$ in the magnetic state. This indicates that quantum criticality in UTe$_2$ is important for the formation of the antiferromagnetic state. This leads to the enhancement of superconductivity under magnetic field close to the critical pressure and to the reentrant behavior of superconductivity along the $c$ axis \cite{Aoki2021, Valiska2021}. However, the transition from the superconducting ground state to the magnetically ordered may be first order and connected to a strong change in the electronic structure. Indication for this is the abrupt increase in the residual resistivity through $p_c$ \cite{Aoki2021, Wilhelm2023}. Direct microscopic evidence for this is given by the change by 7 percent of the 5$f$ count through $p_c$ towards the U$^{4+}$ configuration \cite{Wilhelm2023}. As already mentioned, this change in the electronic structure goes along with a drastic change in the magnetic anisotropy. 

\section{Conclusions}

\label{conclusion}

We have shown that the resistivity in UTe$_2$ depends strongly on the electrical current direction. The measurements with $J \parallel c$ clearly reveal the important energy scales. In particular, at the metamagnetic transtion at $\mu_0 H_m \approx 34.5$~T the resistivity $\rho_c$ strongly decreases for current applied along the $c$ axis, while $\rho_a$ and $\rho_b$ strongly increase (current along $a$ or $b$ axis respectively).
The field dependence of the $A$ coefficient for $J\parallel c$ seems to replicate $\gamma (H)$ better than for the other current directions. 
 The maximum $\Tmaxc$ in the temperature dependence of $\rho_c$ decrease on approaching the critical point of the first order metamagnetic transition line. $\Tmaxc$ coincides with the temperature of the minimum observed in the thermal expansion $\alpha_b$ along the $b$ axis. Under hydrostatic pressure $\Tmaxc$ and $H_m$ decreases up to the critical pressure $p_c$.  The phase diagram in the normal state below $p_c$ scales with $\Tmaxc$, indicating the importance of this low energy scale: the coherence along the $c$ axis governs the low temperature behavior. Superconductivity is observed for all pressures up to $H_m$, \added{which collapses at $p_c$}. At zero pressure the low field superconducting phase lfSC scales with $T_{sc}$ independently of the sample quality, while the high field superconducting phase hfSC is strongly sample dependent. This indicated that these two phases may emerge from different pairing mechanisms \cite{Rosuel2023}. \deleted[id=G]{The lfSC phase is suppressed under pressure and close to $p_c$ only the hfSC phase is observed suggesting that the fluctuations responsible for the hfSC phase at zero pressure drive also superconductivity close to $p_c$.} \added{Accurate determination of the phase diagrams under pressure by thermodynamic measurements under magnetic field along $b$ are strongly required, in particular close to the pressure where the high-temperature superconducting phase occurs. This will elucidate whether the high temperature superconducting phase is the pressure evolution of the field-induced hfSC phase, or not.} A challenge in future will be to precise the field and pressure dependence of the valence and spin correlations and their feedback on superconductivity under pressure.

\section{Acknowledgments}

We thank K. Ishida, Y. Yanase and M.E. Zhitomirsky for fruitful discussions. We received financial support from the French Research Agency ANR within the projects FRESCO
No. ANR-20-CE30-0020 and FETTOM No. ANR-19-CE30-0037. Further finacial support has been provided by the JPSJ programs KAKENHI P22H04933, JP20K20889, JP20H00130, JP20KK0061.  
We acknowledge support of the LNCMI-CNRS, member of the European Magnetic Field Laboratory (EMFL) and from the Laboratorie d'excellence LANEF (ANR-10-LABEX-51-01).

\bibliography{D:/PAPER/UTe2/UTe2}

\begin{thebibliography}{81}
\bibitem [{foo({\natexlab{a}})}]{footnote1}%
  \BibitemOpen
  \bibinfo {note} {Only close to the critical metamagnetic field, the
  temperature dependence seems to be stronger than $T^2$, however, in the
  analysis we forced the fit to be $T^2$, neglecting the initial curvature in
  this plot very close to $\Tc$.}\BibitemShut {Stop}%
\end{thebibliography}

\newpage

\appendix*
\section{Supplementary data}

\label{supplementary_data}

In the Appendix we show additional figures to which we refer in the main text.

\begin{figure}[h]
	\includegraphics[width=0.9\linewidth]{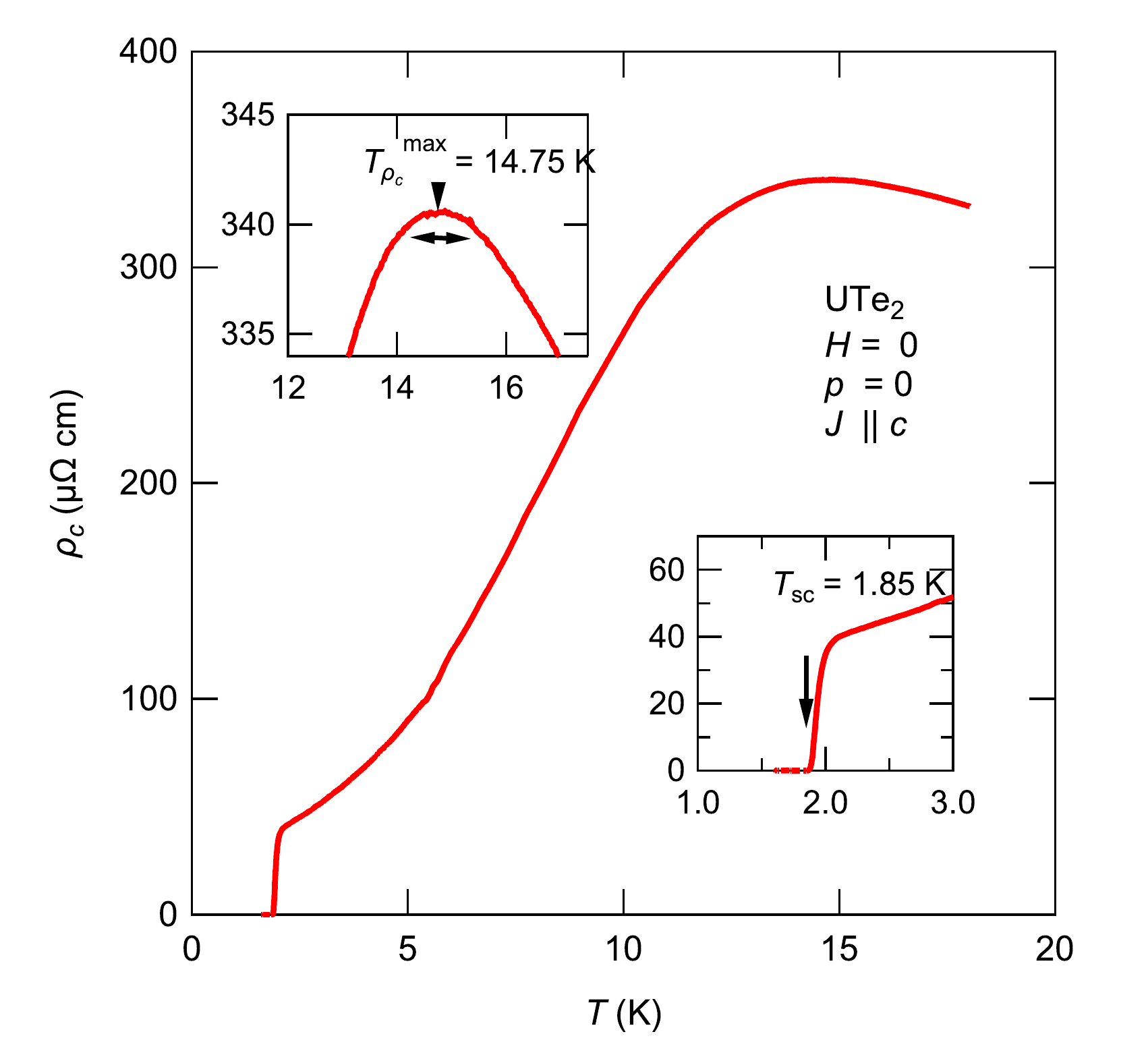}
	\caption{Example for the definition of the superconducting transition temparation $\Tc$ and the maximum $T^\star$ from the resistivity data at $p = 0$ and $h=0$. In the phase diagrams in the following the error bars for $\Tc$ is in all cases smaller than the symbol size and thus we do not plot error bars. We note that the discussion of the transition width from resistivity data alone is difficult as it is not a bulk probe. The upper inset shows a zoom on the anomaly at $T^\star$, the horizontal arrow shows the maximal error of the determination. }
	\label{def_Tsc_Tstar}
\end{figure}

\begin{figure}[h]
	\includegraphics[width=1\linewidth]{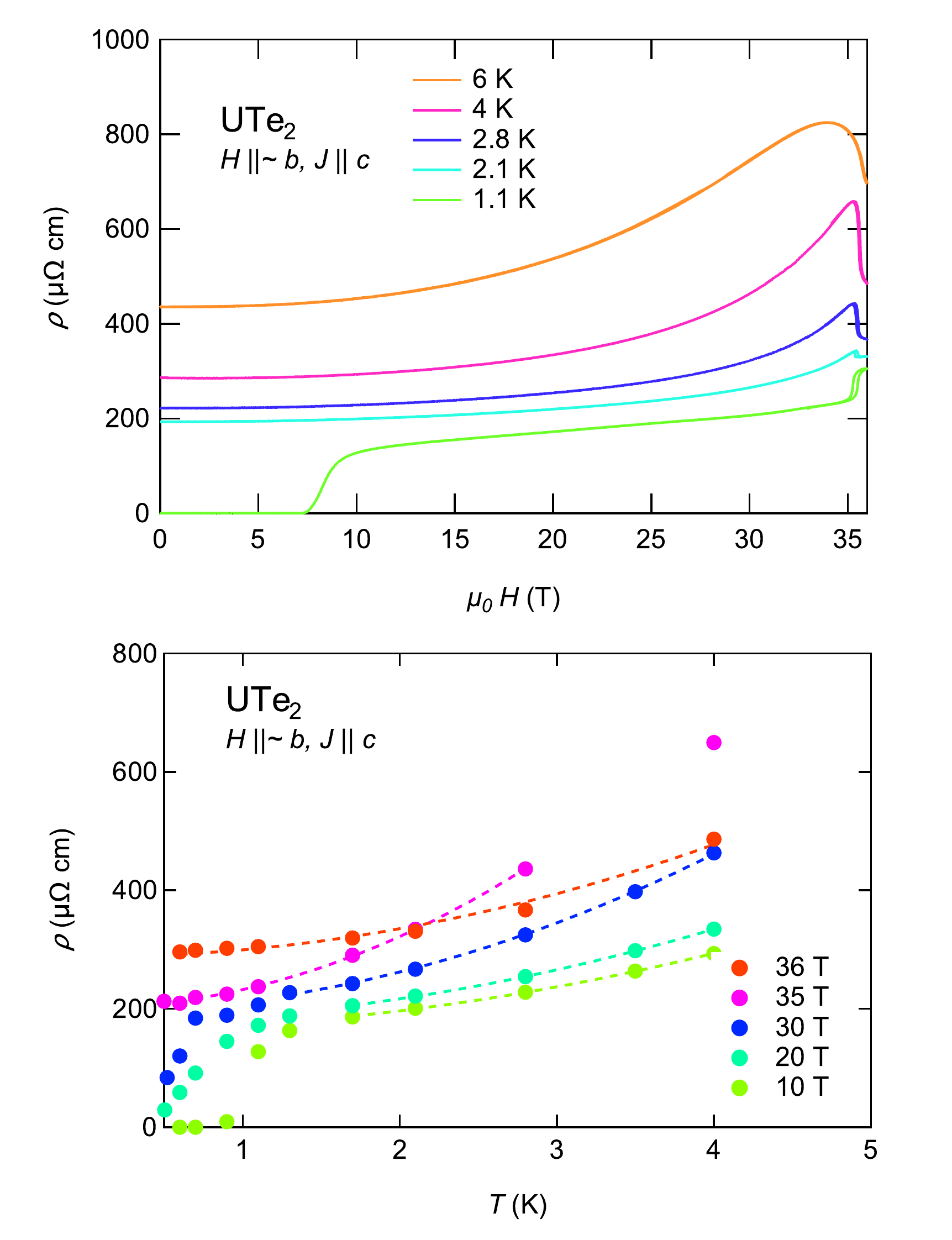}
	\caption{(a) Magnetoresistance $\rho_c$ for field $H \parallel \sim b$ and current along $c$ measured on sample S5 for different temperatures $T$ between 1 and 6~K.  The metamagnetic transition appears at $H_m = 35.4$~T in the field up sweeps. The hysteresis of the metamagnetic transition is about 0.4~T. While for $T=1.1$~K the jump at $H_m$ is positive, it is negative and very tiny at 2.1~K in the when the sample is in the normal state for all fields.The lower panel (b) shows the temperature dependence extracted from the field dependent measurements. Fermi liquid fits (dashed lines) are performed in the temperature range $T<4$~K. }
	\label{Annex_rho_c}
\end{figure}

\begin{figure}[h]
	\includegraphics[width=0.9\linewidth]{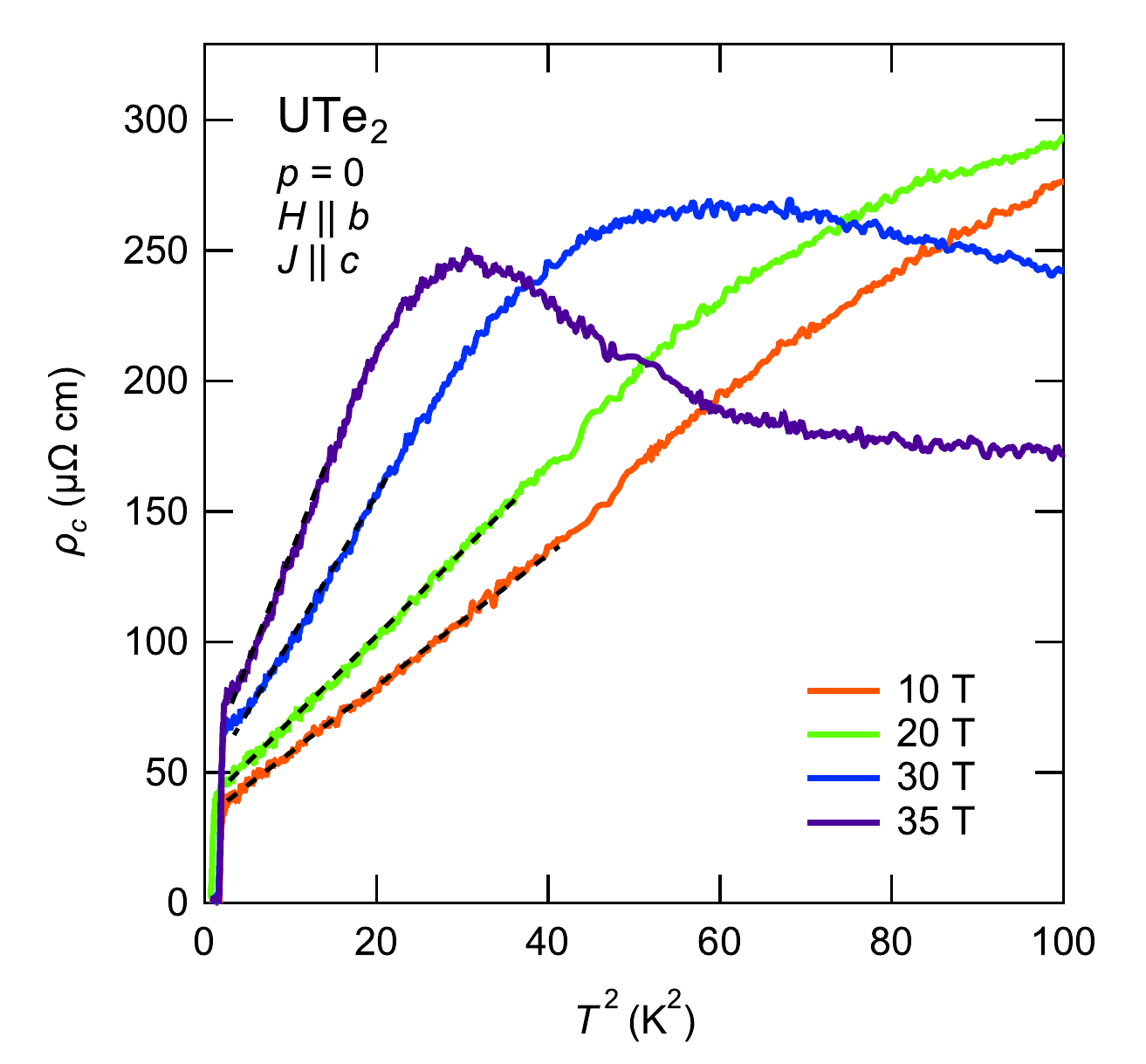}
	\caption{(a) Temperature dependence of the resistivity $\rho_c$ as a function of $T^2$ for different magnetic fields up to 34 T. The dashed-lines present a fit with $\rho_c = \rho_0 + A T^2$. The range of the $T^2$ regime decreases with increasing  magnetic field. Close to the metamagnetic field a fit with an free exponent $n$ (Figure \ref{Annex_rho_c}(a)) would result in an exponent $n = 3$, but we in the analysis of the Fermi liquid $A$ coefficient, we fixed the exponent to be $n = 2$.   }
	\label{rho_vs_T2}
\end{figure}



\begin{figure}[h]
	\includegraphics[width=1\linewidth]{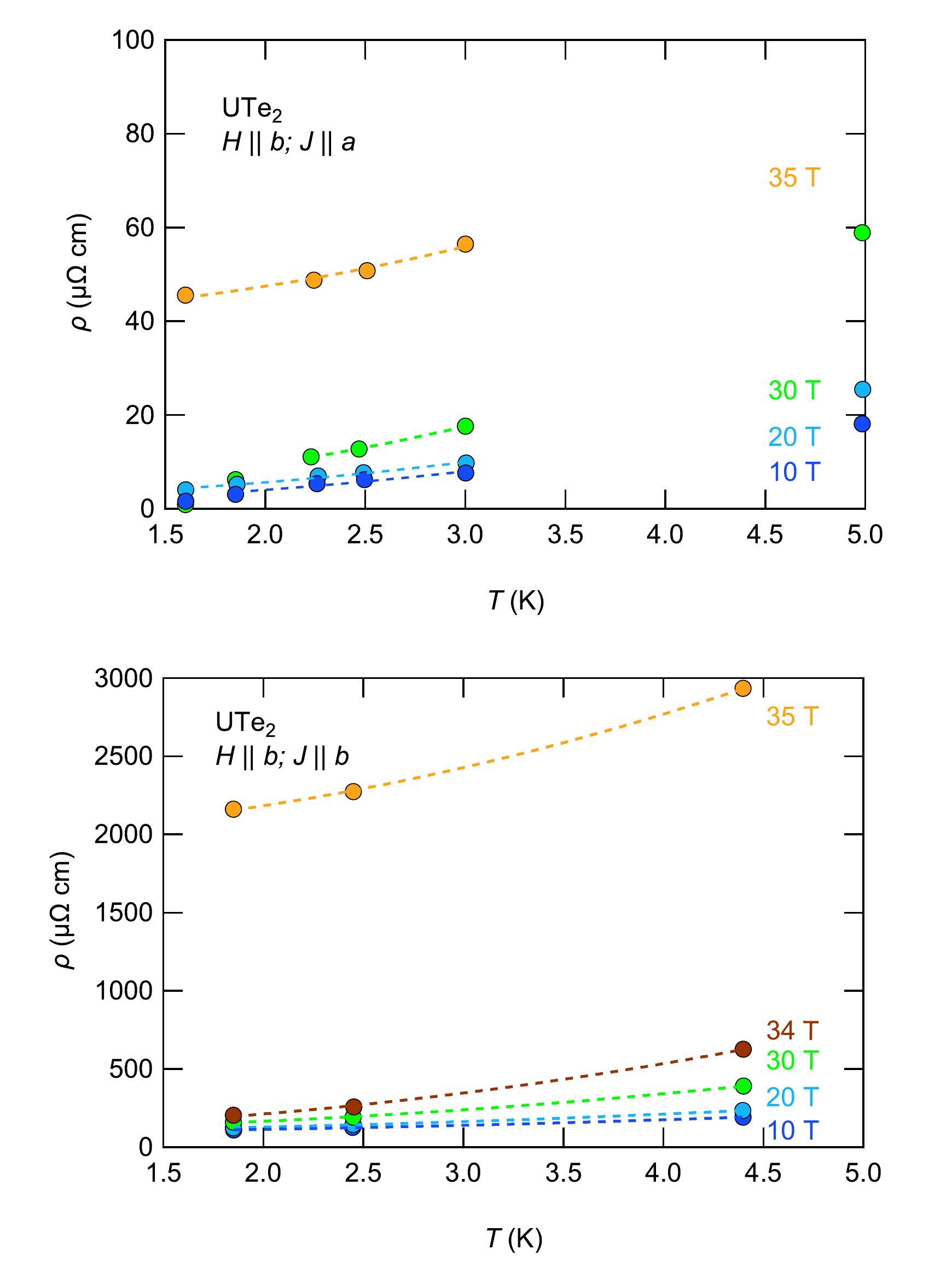}
	\caption{(a) Temperature dependence of the resistivity for different magnetic fields reconstructed from the magnetoresistance measurements show for field $H \parallel \sim b$ measured on sample S3 for $J \parallel a$ (upper panel) and on sample S4 for $J\parallel b$ (lower panel). $T^2$ fits are indicated by the dotted lines, due to the limited temperatures error bars for the $A$ coefficient shown in Fig.~\ref{Fig05} are rather large.}
	\label{Annex_rho_Jb_and_Ja}
\end{figure}

\begin{figure}[h]
	\includegraphics[width=1\linewidth]{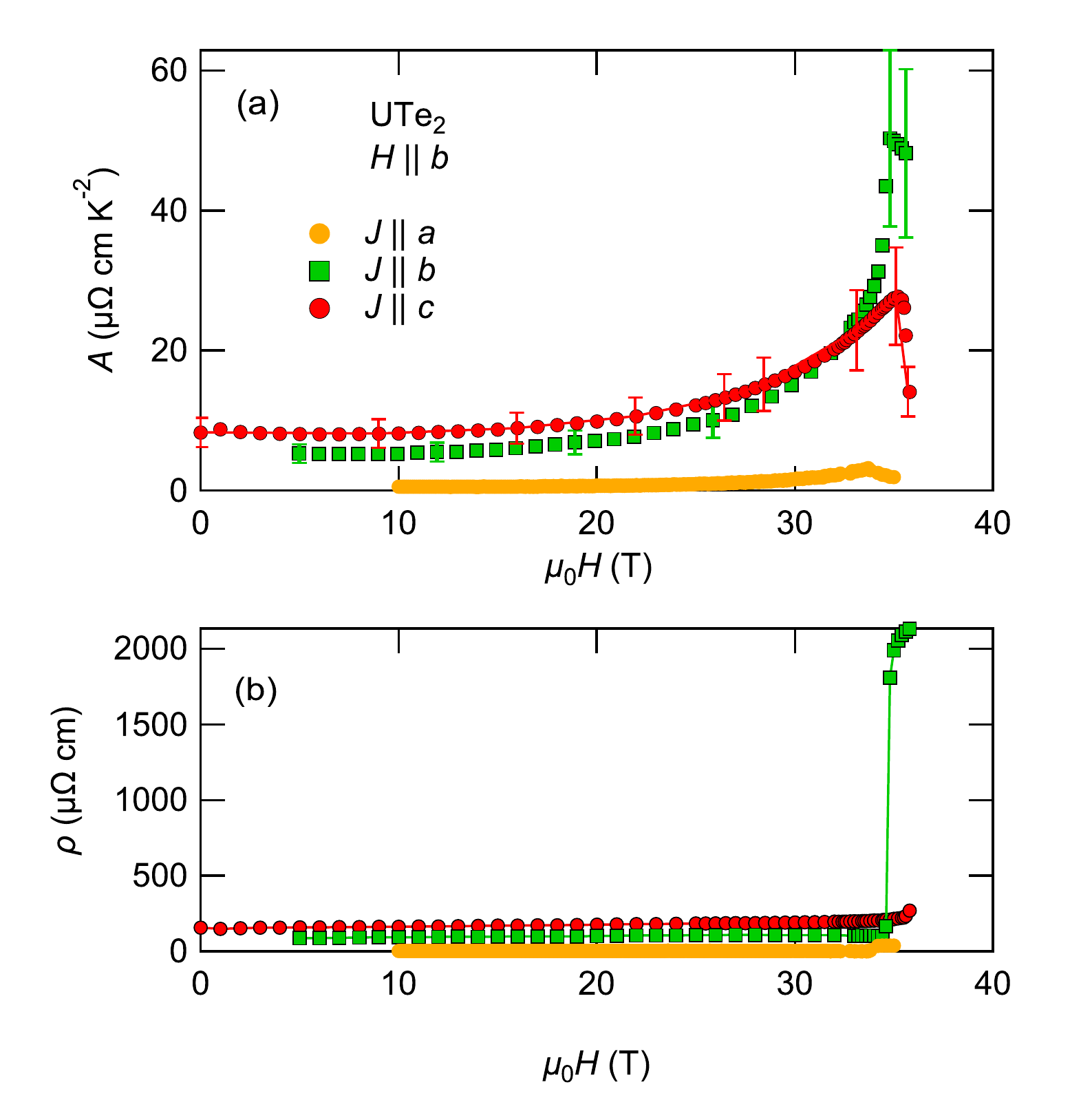}
	\caption{\added{(a) Fermi-liquid coefficient $A$ as a function of the magnetic field $H \parallel b$ with current $J$ applied along $a$, $b$, and $c$ axis. (b) Field dependence of the residual resistivity $rho_0$ for the different current directions. } }
	\label{Annex_A_coef_all_directions_absolute}
\end{figure}

\begin{figure}[h]
	\includegraphics[width=1\linewidth]{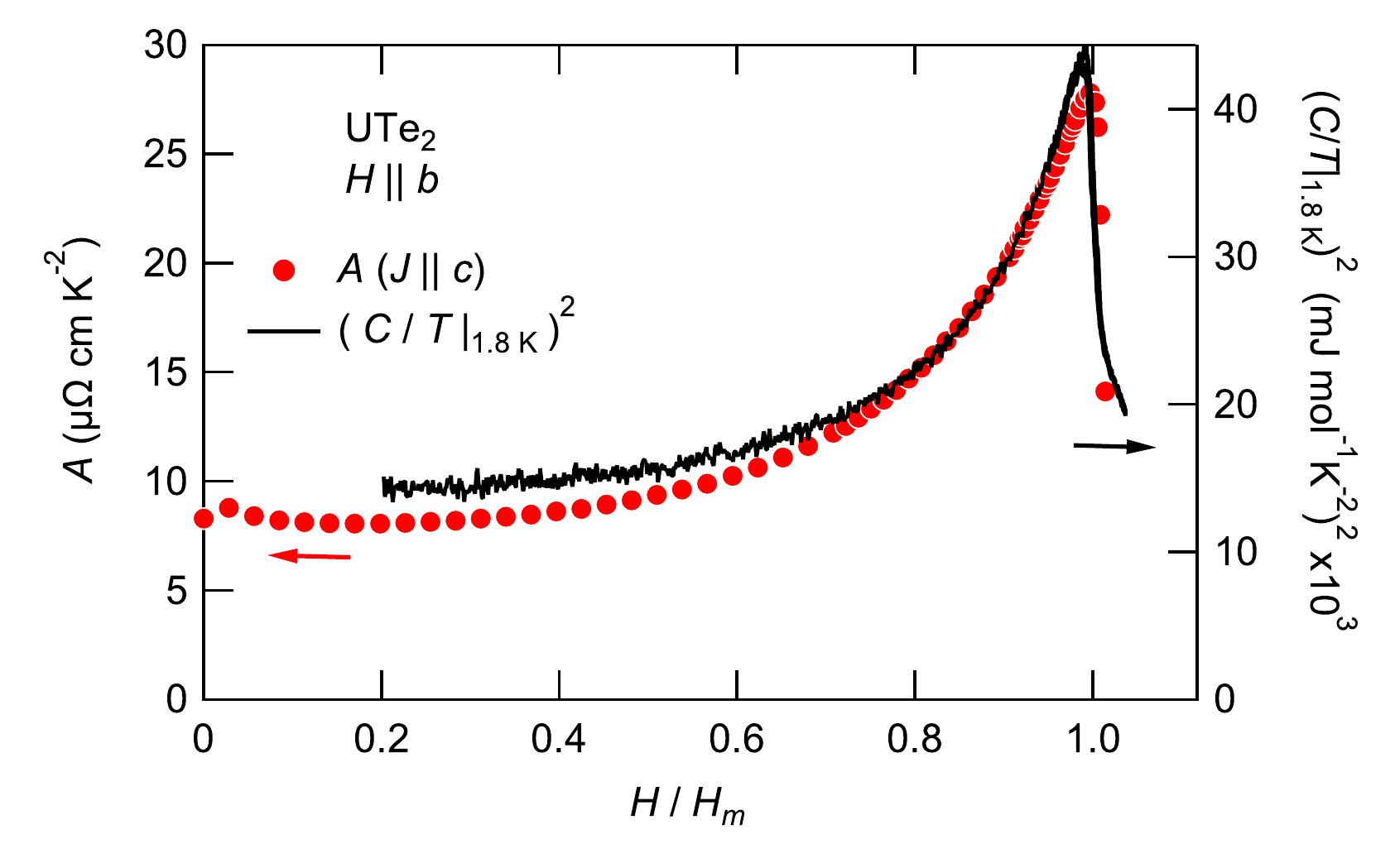}
	\caption{Field dependence of the $A$ coefficient determined from the resistivity measurements with $J \parallel c$ and $(C/T)^2$ measured at 1.8~K as a function of magnetic field normalized to the metamagnetic transition field $H_m$. Both quantities scale surprisingly well. Especially the very sharp drop at the metamagnetic transition is only observed in the measurement of the $A$ coefficient with current applied along the $c$ axis, while for current in the other directions, the enhancement of the Fermi-liquid $A$ coefficient occurs to be more symmetric around $H_m$. }
	\label{Annex_A_gamma}
\end{figure}

 
\begin{figure}[h]
	\includegraphics[width=1\linewidth]{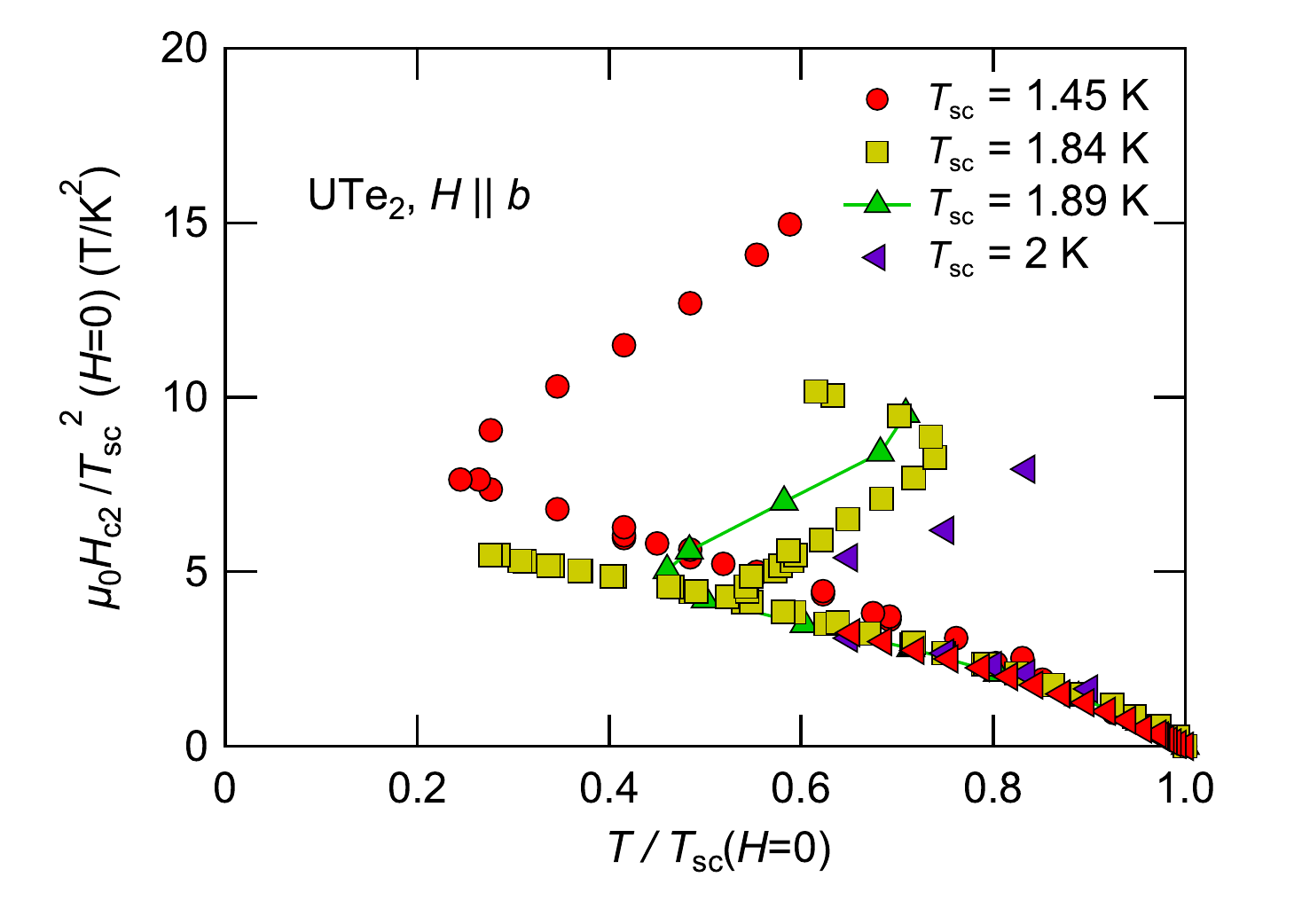}
	\caption{Upper critical field $\Hc$ normalized by $\Tc^2$ versus  $T/\Tc$ for the different samples. This scaling would correspond to an purely orbital limited upper critical field. However the scaling of the critical field only by $\Tc$ shown in Fig.~\ref{comparisonpd} seems to be better. }
	\label{Annex_comparison_PD}
\end{figure}

\begin{figure}[h]
	\includegraphics[width=1\linewidth]{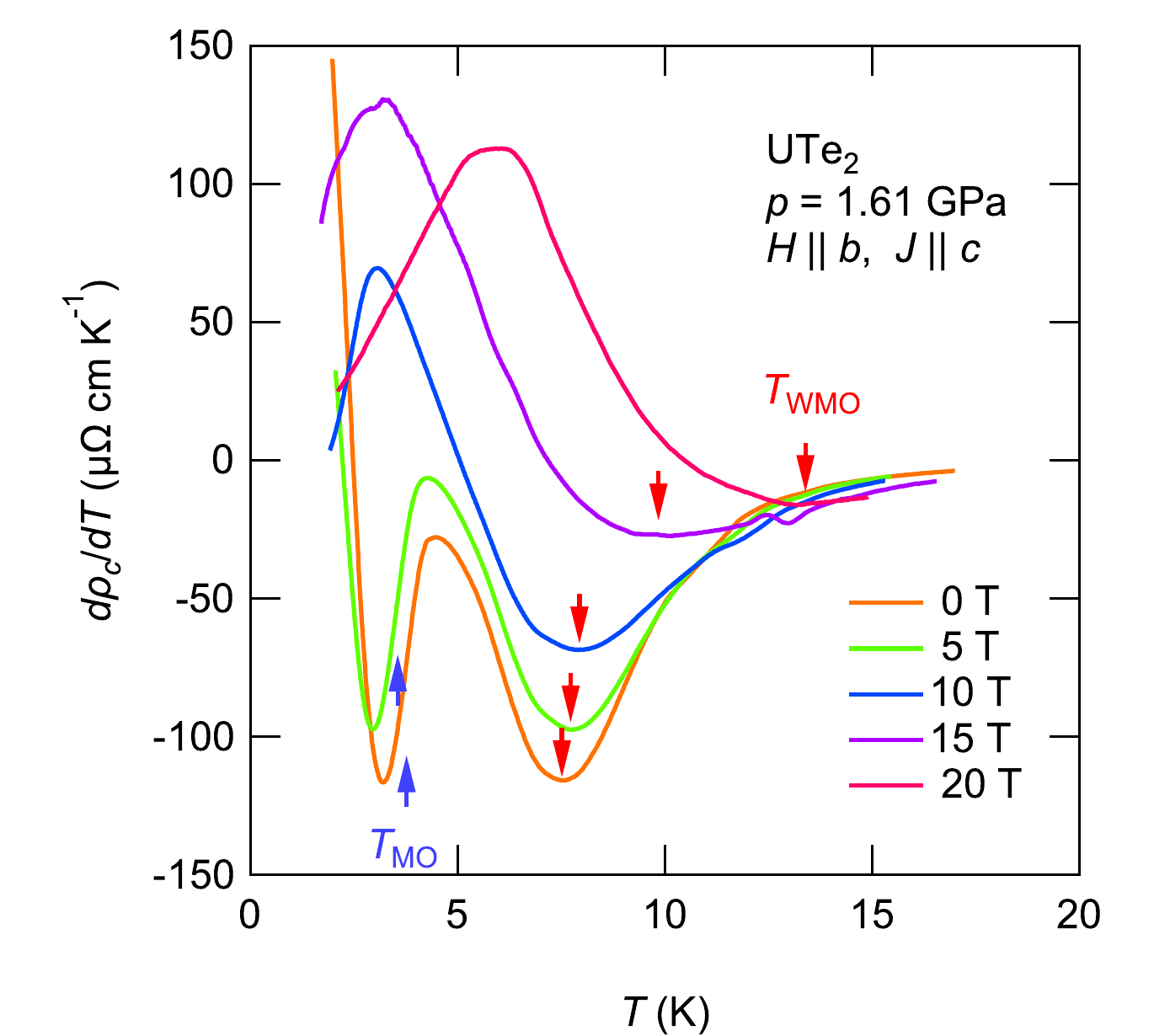}
	\caption{Temperature derivative $d\rho_c/dT$ at $p=1.61$~GPa for different magnetic fields. The arros indicate the magnetic transition temperature $\TMO$ and the temperature of the crossover $\TWMO$.}
	\label{Annex_dif_P5}
\end{figure}

\begin{figure}[h]
	\includegraphics[width=1\linewidth]{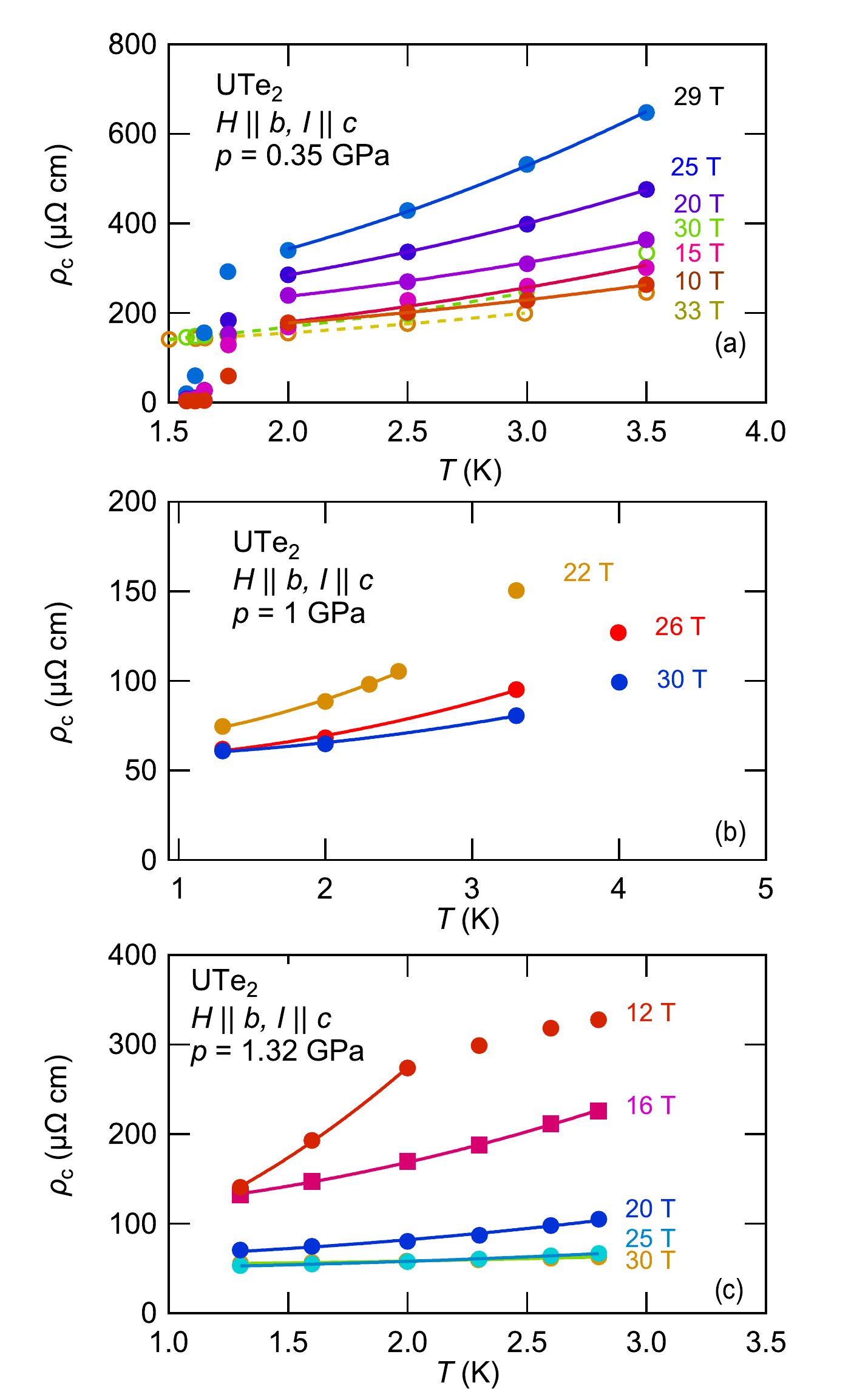}
	\caption{Temperature dependence of the resistivity reconstructed from the measured magnetoresistance for different magnetic field for different pressures, (a) 0.35~GPa, (b) 1~GPa, and (c) 1.32~GPa. }
	\label{Annex_rho_T2_pressure}
\end{figure}

\begin{figure}[h]
	\includegraphics[width=1\linewidth]{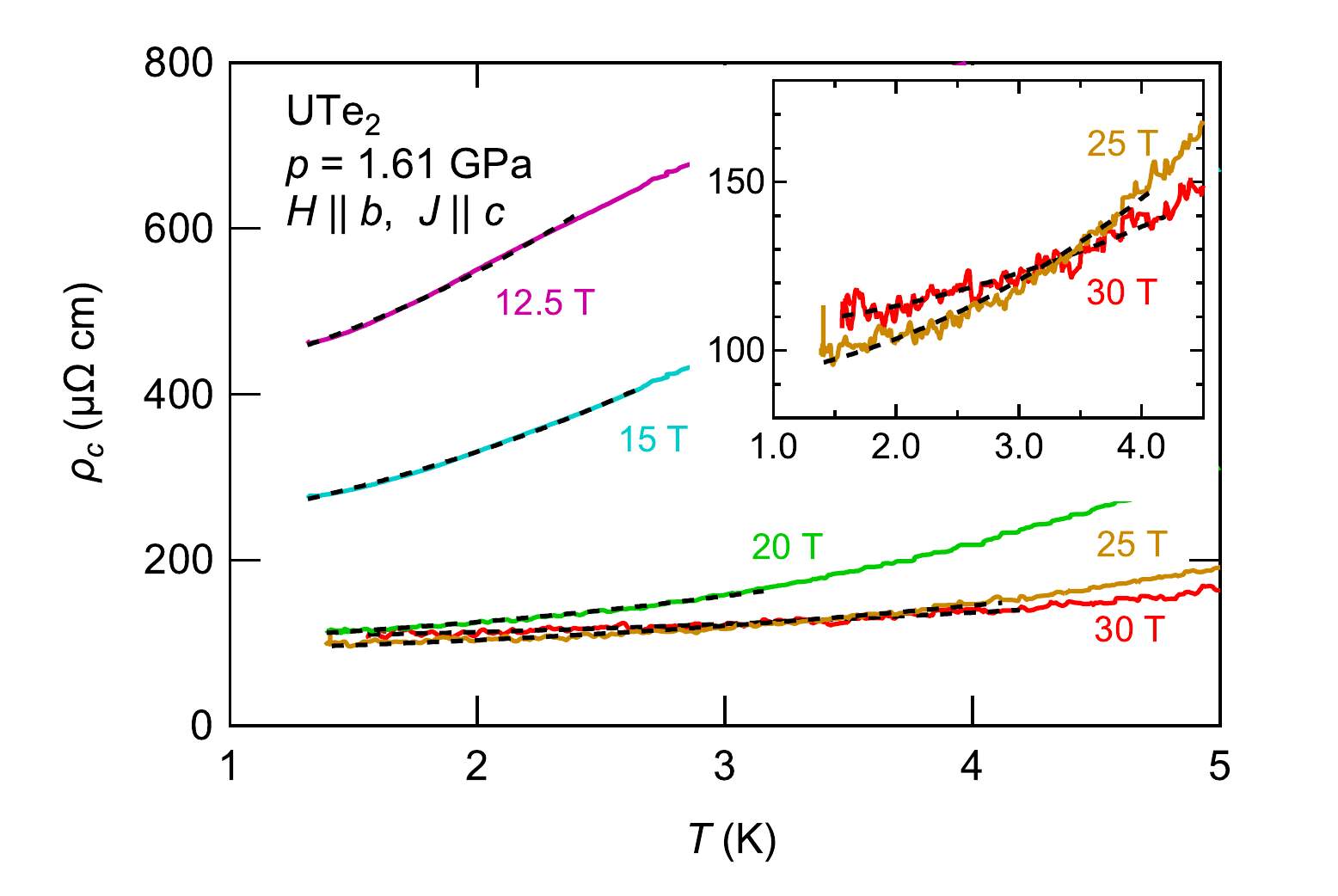}
	\caption{Temperature dependence of the resistiviy $\rho_c$ at 1.61~GPa as a function of temperature up to 5~K for different magnetic fields in the polarized magnetic regime above the critical field $H_c$ of  the antiferromagnetic state. The dashed lines indicate fits with a Fermi liquid temperature dependence $\rho_c = \rho_0 + AT^2$. The temperature range of the Fermi-liquid range decreases strongly on appoaching $H_c$. \added{The inset shows the data at 25~T and 30~T in an expanded view.}  }
	\label{Annex_rho_T2_P5}
\end{figure}

\end{document}